\documentstyle[prl,aps,multicol,eqsecnum]{revtex}

\input epsf

\preprint{Submitted to {\it Physical Review A}}

\draft

\author{Howard~Barnum,$^{1,2}$ \thanks{hbarnum@tangelo.phys.unm.edu}, M.~A.~Nielsen \thanks{%
mnielsen@tangelo.phys.unm.edu}$^{1,2}$,
and
Benjamin
Schumacher$^3$
\thanks{schumacb@kenyon.edu}}
\address{$^1$ Institute for Theoretical Physics, \\ University of
 California,
Santa
Barbara
CA
93106-4030}
\address{$^2$ Center for Advanced Studies,
  Department of Physics and Astronomy,\\
 University of New Mexico, Albuquerque, NM 87131-1156}
\address{$^3$ Department of Physics, Kenyon College, Gambier, OH 43022}
\title{Information transmission through a noisy quantum channel}
\date{\today}

\begin{document}

\pagestyle{plain}
\pagenumbering{arabic}

\maketitle

\begin{abstract}
Noisy quantum channels may be used in many information carrying
applications. We show that different applications may result in different
channel capacities. Upper bounds on several of these capacities are
proved. These bounds are based on the {\em coherent information}, which
plays a role in quantum information theory analogous to that played by
the mutual information in classical information theory. Many new properties
of the coherent information and entanglement fidelity 
are proved. Two non-classical features of the coherent information are
demonstrated: the failure of subadditivity, and the failure of
the pipelining inequality. Both properties
arise as a consequence of quantum entanglement, and give
quantum information new features not found in classical information theory.
The problem of a noisy quantum channel with a classical observer
measuring the environment is introduced,
and bounds on the corresponding channel capacity
proved. These bounds are always greater than for the unobserved channel. We
conclude with a summary of open problems.
\end{abstract}


\begin{multicols}{2}[]
\narrowtext

\section{Introduction}

A central result of Shannon's classical theory of information 
\cite{Shannon48a,Shannon49a,Cover91a} is
the {\em noisy channel coding theorem}. This result provides an
{\em effective procedure} for determining the
{\em capacity} of a noisy channel - the maximum
rate at which classical information can be reliably transmitted
through the channel. There has been much recent work on quantum
analogues of this result
\cite{Schumacher96a,Schumacher96b,Bennett96a,Lloyd97a,Bennett97a}.  

This paper has two central purposes. The first purpose is to develop
general techniques for proving upper bounds on the capacity of a noisy
quantum channel, which are applied to several different
classes of quantum noisy channel problems. Second, we point out some
essentially new features that quantum mechanics introduces into the noisy
channel problem. 

The paper is organized as follows. In section \ref{sect: channels} we
give a basic introduction to the problem of the noisy quantum channel,
and explain the key concepts. Section \ref{sect: qops} reviews the
{\em quantum operations} formalism that is used throughout the
paper to describe a noisy quantum channel, and section
\ref{sect: entropy exchange} reviews the concept of the {\em entropy exchange}
associated with a quantum operation. Section \ref{sect: classical} shows how
the classical noisy channel coding theorem can be put into the quantum
language, and explains why the capacities that arise in this context
are not useful for applications such as quantum computing and
teleportation. Section
\ref{sect: entanglement fidelity} discusses the {\em entanglement fidelity},
which is the measure we use to quantify how well a state and its entanglement
are transmitted through a noisy quantum channel. Section
\ref{sect: coherent information} discusses the {\em coherent information}
introduced in \cite{Schumacher96b} as an analogue to the concept of
{\em mutual information} in classical information theory. Many new results
about the coherent information are proved, and we show that quantum
entanglement allows the coherent information to have properties which have no
classical analogue. These properties are critical to understanding
what is essentially quantum about the quantum noisy channel coding problem.
Section \ref{sect: noisy coding revisited} brings us back to noisy channel
coding, and formally sets
up the class of noisy channel coding problems we
consider. Section \ref{sect: upper bounds} proves a variety of
upper bounds on the capacity of a noisy quantum channel, depending on
what class of coding schemes one is willing to allow. This is followed
in section \ref{sect: discussion} by a discussion of the achievability
of these upper bounds and of earlier work
on channel capacity.
Section \ref{sect: observed channel} formulates the
new problem of a noisy quantum channel with measurement, allowing
classical information about the environment to be obtained by measurement,
and then used during the decoding process. Upper bounds on the corresponding
channel capacity are proved. Finally, section
\ref{sect: conc} concludes with a summary of our results,
a discussion of the new features which quantum mechanics adds to the
problem of the noisy channel, and suggestions for further research.

\section{Noisy Channel Coding}
\label{sect: channels}

\begin{figure}
\epsfxsize 3.4in
\epsfbox{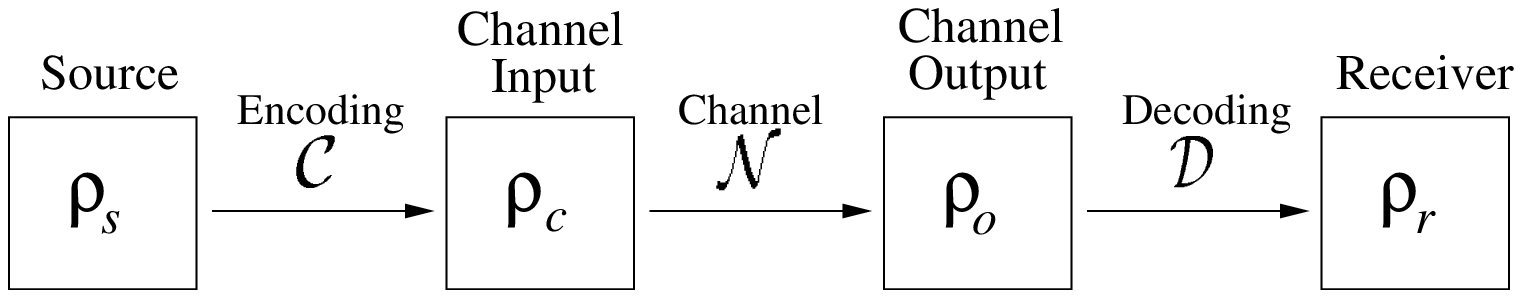}
\caption{} \label{fig: channel0}
The noisy quantum channel, together with encodings and decodings.
\end{figure}

The problem of noisy channel coding will be outlined in this section.
Precise definitions of the concepts used will
be given in later sections. The procedure is illustrated in
figure \ref{fig: channel0}.

There is a {\em quantum source} emitting unknown quantum states, which
we wish to transmit through the channel to some receiver. Unfortunately,
the channel is  usually subject to noise, which prevents
it from transmitting states with high fidelity. For example, an optical
fiber suffers losses during transmission. Another important example
of a noisy quantum channel is the memory of a quantum computer. There the
idea is to transmit quantum states {\em in time}. The effect of transmitting
a state from time $t_1$ to $t_2$ can be described as a noisy
quantum channel. Quantum teleportation \cite{Bennett93a}
can also be described as a noisy quantum channel whenever there
are imperfections in the teleportation process \cite{Bennett96a,Nielsen96b}.
 
The idea of noisy channel coding is to encode the quantum state emitted by
the source, $\rho_s$, which one wishes to transmit, 
using some {\em encoding operation}, which we denote
${\cal C}$. The encoded state is then sent through the channel, whose
operation we denote by ${\cal N}$. The output state of the channel
is then {\em decoded} using some {\em decoding operation}, ${\cal D}$.
The objective is for the decoded state to match with high fidelity the
state emitted by the source.  As in the classical theory, we 
consider the fidelity of large blocks of material produced by 
repeated emission from the source, and allow the encoding and decoding
to operate on these blocks.  A channel is said to transmit a
source reliably if a sequence of block-coding and block-decoding
procedures can be found that approaches perfect fidelity in the
limit of large block size.

What then is the {\em capacity} of such a channel - the highest rate at
which information can be reliably transmitted through the channel? The
goal of a {\em channel capacity theorem} is to provide a procedure
to answer this question. This procedure must be an {\em effective
procedure}, that is, an explicit algorithm to evaluate the
channel capacity. 
Such a theorem comes in two parts. One part proves an upper bound on
the rate at which information can be reliably transmitted through the channel.
The other part demonstrates that there are coding and decoding schemes which
attain this bound, which is therefore the channel capacity. 
We  do not prove such a channel capacity theorem in this paper. We do,
however, derive bounds
on the rate at which information can be sent through a noisy quantum channel.

\section{Quantum Operations}
\label{sect: qops}

What is a quantum noisy channel, and how can it be described
mathematically? This section reviews the formalism
of quantum operations, which is used to describe noisy channels. Previous
papers on the noisy channel problem
\cite{Schumacher96a,Schumacher96b,Bennett96a,Lloyd97a,Bennett97a}
have used apparently different formalisms to describe the noisy channel.
In fact, all the formalisms can be shown to be equivalent, as we shall see
in this section. Historically, quantum operations have also sometimes
been known as {\em completely positive maps} or
{\em superscattering operators}. The motivation in all cases has been
to describe general state changes in quantum mechanics.

A simple example of a state change in quantum mechanics is
the unitary evolution experienced by a closed quantum system. 
The final state of the system is related to the initial
state by a unitary transformation $U$,
\begin{eqnarray} \label{eqtn: unitary operation}
\rho \rightarrow {\cal E}(\rho) = U \rho U^{\dagger}\;. \end{eqnarray}
Although all closed quantum systems are described by unitary
evolutions, in accordance with Schr\"odinger's equation, more general
state changes are possible for open quantum systems, such as noisy quantum
channels.

How does one describe a general state change in 
quantum mechanics? The answer to this question is provided by the 
quantum operations formalism.  This formalism is described in 
detail by Kraus \cite{Kraus83a} (see also Hellwig and Kraus \cite{Hellwig70})
and is given short but detailed 
reviews in Choi \cite{Choi75a} and in the Appendix to \cite{Schumacher96a}. 
In this formalism there is an {\em input state\/} and an {\em output 
state}, which are connected by a map
\begin{eqnarray} \label{eqtn: general evolution}
\rho \rightarrow 
\frac{{\cal E}(\rho)}{\mbox{tr}\bigl({\cal E}(\rho)\bigr)}\;.
\end{eqnarray}
This map is a {\em quantum operation\/}, ${\cal E}$, 
a linear, trace-decreasing map that preserves positivity.  The trace 
in the denominator is included in order to preserve the trace condition,
$\mbox{tr}(\rho) = 1$.

The most general form for ${\cal E}$ that is physically reasonable 
(in addition to being linear and trace-decreasing and preserving
positivity, a physically reasonable ${\cal E}$ must satisfy an 
additional property called complete positivity),
can be shown to be \cite{Kraus83a}
\begin{eqnarray} \label{eqtn: gen op}
{\cal E}(\rho) = \sum_i A_i \rho A_i^{\dagger}\;. \end{eqnarray}
The system operators $A_i$, which must satisfy \linebreak
$\sum_i A_i^\dagger A_i\le I$, 
completely specify the quantum operation.  In the particular case
of a unitary transformation, there is only one term in the
sum, $A_1 = U$, 
leaving us with the transformation~(\ref{eqtn: unitary operation}).

A class of operations that is of particular interest are the
{\em trace-preserving} or {\em non-selective} operations. Physically,
these arise in situations where the system is coupled to some environment
which is not under observation; the effect of the evolution is averaged
over all possible outcomes of the interaction with the environment.
Trace-preserving operations are defined by the requirement that
\begin{eqnarray}
\sum_i A_i^{\dagger} A_i = I. \end{eqnarray}
This is equivalent to requiring that for all density operators $\rho$,
\begin{eqnarray}
\mbox{tr}({\cal E}(\rho)) = 1, \end{eqnarray}
explaining the nomenclature ``trace-preserving''. Notice that this means
the evolution equation (\ref{eqtn: general evolution}) reduces to the simpler
form
\begin{eqnarray}
\rho \rightarrow {\cal E}(\rho), \end{eqnarray}
when ${\cal E}$ is trace-preserving.

The following {\em representation theorem} is proved in \cite{Kraus83a}, \cite{Choi75a}, and
\cite{Schumacher96a}. It shows the connection between trace-preserving
quantum operations and systems interacting unitarily with an environment, and
thus provides part of the justification for the physical interpretation
of trace-preserving quantum operations described above.

{\em Theorem} (representation theorem for trace-preserving quantum operations):

Suppose ${\cal E}$ is a trace-preserving quantum operation on a system
with a $d$-dimensional state space. Then it is
 possible to construct an ``environment'' $E$ of at most $d^2$ dimensions,
such that the system plus environment are initially uncorrelated,
the environment is initially in a pure state $\sigma = |s\rangle \langle s|$
and there exists a unitary evolution $U$ on system and environment such that
\begin{eqnarray}
{\cal E}(\rho) = \mbox{tr}_E(U (\rho \otimes \sigma) U^{\dagger}).
\end{eqnarray}
Here and elsewhere in the paper a subscript on a trace indicates a 
partial trace over the corresponding system ($E$ in this case).

Conversely, given any initially uncorrelated environment $\sigma$ 
(possibly of more
than $d^2$ dimensions, and initially impure), a unitary interaction
$U$ between the system and the environment gives rise to a trace-preserving
quantum operation,
\begin{eqnarray}
{\cal E}(\rho) = \mbox{tr}_E(U (\rho \otimes \sigma) U^{\dagger}).
\end{eqnarray}

This theorem tells us that any trace-preserving quantum operation can always
be {\em mocked up} as a unitary evolution by adding an environment
with which the system can interact unitarily. Conversely, it tells
us that any such unitary interaction with an initially uncorrelated
environment gives rise to a trace-preserving quantum operation. Both
these facts are useful in what follows. The picture we have of a quantum
operation is neatly summarized by the following diagram.

\begin{figure}
\begin{center}
\unitlength 1cm
\begin{picture}(8,5)(0,0)
\put(2,3){\framebox(1,1){$Q$}}
\put(3,3.5){\vector(1,0){3}}
\put(6,3){\framebox(1,1){$Q'$}}
%
\put(1.25,3.5){\makebox(0,0){$\rho^{Q}$}}
%
\put(4,0.5){\framebox(1,1){$E$}}
\put(4.4,1.7){\vector(0,1){1.6}}
\put(4.6,3.3){\vector(0,-1){1.6}}
\put(4.8,2.5){\makebox(0,0)[cl]{$U^{QE}$}}
\end{picture}
\caption{} \label{fig: quantum operation}
\end{center}
\end{figure}

Here, $Q$ denotes the state of the system before the interaction with
the environment, and $Q'$ the state of the system after the interaction.
Unless stated otherwise we follow the convention that $Q$ and $Q'$
are $d$-dimensional.
The environment system $E$ and the operator $U^{QE}$ might be chosen to be
the actual physical environment and its interaction with $Q$, but this is
not necessary. The only thing that matters for the description of noisy
channels is the dynamics of $Q$.
For any given quantum operation ${\cal E}$ there are 
many possible representations of ${\cal E}$ in terms
of environments $E$ and interactions $U^{QE}$. We always assume that
the initial state of $E$ is a {\em pure state}, and regard $E$
as a mathematical artifice. Of course, the actual physical environment,
$E_A$, may be initially impure, but the above
representation theorem shows that for the purposes of describing the dynamics
of $Q$, it can be replaced by an ``environment'', $E$, which is initially
pure and gives rise to exactly the same dynamics. In what follows it is
this latter $E$ that is  most useful.

Shannon's classical noisy coding theorem is proved for {\em discrete
memoryless channels}. Discrete means that the channel only has a finite
number of input and output states. By analogy we define a discrete
quantum channel to be one which has a finite number of Hilbert space
dimensions. In the classical case, memoryless means that the output of the
channel is independent of the past, conditioned on knowing the state of
the source. Quantum mechanically we take this to mean
that the output of the channel is completely determined by the encoded
state of the source, and is not affected by the previous history of the source.

Phrased in the language of quantum operations, we assume that there
is a quantum operation, ${\cal N}$, describing the dynamics of the channel.
The input $\rho_i$ of the channel is related to the output $\rho_o$
by the equation
\begin{eqnarray}
\rho_i \rightarrow \rho_o = {\cal N}(\rho_i). \end{eqnarray}
For the majority of this paper we assume, as in the
previous equation, that the operation describing the action of the
channel is trace-preserving. This corresponds to the physical assumption that
no classical information about the state of the system or its environment
is obtained by an external classical observer. All previous work on
noisy channel coding with the exception of \cite{Nielsen97a}
has assumed that this is the case, and we do so for the majority
of the paper. In section \ref{sect: observed channel} we
consider the case of a noisy channel which is being observed by some
classical observer.

In addition to the environment, $E$, it is also extremely useful to
introduce a {\em reference system}, $R$, in the following way. One might
imagine that the system $Q$ is initially part of a larger system, $RQ$,
and that the total is in a pure state, $|\psi^{RQ}\rangle$, satisfying
\begin{eqnarray}
\rho^Q = \mbox{tr}_R (|\psi^{RQ}\rangle \langle \psi^{RQ}|).
\end{eqnarray}
Such a state $|\psi^{RQ}\rangle$ is called a {\em purification} of
$\rho^Q$, and it can be shown \cite{Peres93a} that such a system $R$ and
purifications $|\psi^{RQ}\rangle$ always exist. From our point of view
$R$ is introduced simply as a mathematical device to purify the initial
state. The joint system $RQ$ evolves according to the dynamics
${\cal I}_R \otimes {\cal E}$ given by
\begin{eqnarray}
\rho^{R'Q'} = ({\cal I}_R \otimes {\cal E})(\rho^{RQ}). \end{eqnarray}

The overall picture we have of a trace-preserving quantum operation thus looks
like:
\begin{figure}
\begin{center}
\unitlength 1cm
\begin{picture}(8,7)(0,0)
\put(2,3){\framebox(1,1){$Q$}}
\put(3,3.5){\vector(1,0){3}}
\put(6,3){\framebox(1,1){$Q'$}}
%
\put(2,5){\framebox(1,1){$R$}}
\put(2.4,4.10){\line(0,1){.05}} \put(2.6,4.10){\line(0,1){.05}}
\put(2.4,4.25){\line(0,1){.05}}  \put(2.6,4.25){\line(0,1){.05}}
\put(2.4,4.40){\line(0,1){.05}}  \put(2.6,4.40){\line(0,1){.05}}
\put(2.4,4.55){\line(0,1){.05}}  \put(2.6,4.55){\line(0,1){.05}}
\put(2.4,4.70){\line(0,1){.05}}  \put(2.6,4.70){\line(0,1){.05}}
\put(2.4,4.85){\line(0,1){.05}}  \put(2.6,4.85){\line(0,1){.05}}
\put(1,4.5){\makebox(0,0){$\left | \Psi^{RQ} \right \rangle$}}
%
\put(4,0.5){\framebox(1,1){$E$}}
\put(4.4,1.7){\vector(0,1){1.6}}
\put(4.6,3.3){\vector(0,-1){1.6}}
\put(4.8,2.5){\makebox(0,0)[cl]{$U^{QE}$}}
\end{picture}
\end{center}
\caption{} \label{fig: quantum operation 2}
\end{figure}

The picture we have described thus far applies only to {\em trace-preserving}
quantum operations.
Later in the paper we will also be interested in quantum operations which
are not trace-preserving. That is, they do not satisfy the
relation $\sum_i A_i^{\dagger} A_i = I$, and thus $\mbox{tr}({\cal E}(\rho))
\neq 1$ in general. Such quantum operations arise in the theory of
{\em generalized measurements}. To each outcome, $m$, of a measurement
there is an associated quantum operation, ${\cal E}_m$, with an
operator-sum representation,
\begin{eqnarray}
{\cal E}_m(\rho) = \sum_i A_{mi} \rho A_{mi}^{\dagger}. \end{eqnarray}
The probability of obtaining outcome $m$ is postulated to be
\begin{eqnarray}
\mbox{Pr}(m) = \mbox{tr}({\cal E}_m(\rho)) =
	 \mbox{tr}(\sum_i A_{mi}^{\dagger} A_{mi} \rho). \end{eqnarray}
The completeness relation for probabilities \linebreak $\sum_m \mbox{Pr}(m) = 1$
is equivalent to the completeness relation for the operators appearing
in the operator-sum representations
\begin{eqnarray}
\sum_{mi} A_{mi}^{\dagger} A_{mi} = I. \end{eqnarray}
Thus for each $m$,
\begin{eqnarray}
\sum_i A_{mi}^{\dagger} A_{mi} \leq I. \end{eqnarray}
As an aside, it is interesting to note that the formulation of quantum  measurement
based on the projection postulate \cite{Cohen-Tannoudji77a,Hughes89,Luders51},
taught in most classes on quantum mechanics, is a special case of
the quantum operations formalism, obtainable by using a single projector
$A_m = P_m$ in the operator-sum representation for ${\cal E}_m$. The
formalism of positive operator valued measures (POVMs) \cite{Peres93a}
is also related to the generalized measurements formalism:
$E_m \equiv \sum_i A_{mi}^{\dagger} A_{mi}$ 
are the elements of the POVM which is measured.

A result analogous to the earlier representation theorem for trace-preserving
quantum operations can be proved for general operations.

{\em Theorem} (general representation theorem for operations)

Suppose ${\cal E}$ is a general quantum operation. Then it is possible to find
an environment, $E$, initially in a pure state
$\sigma = |s\rangle \langle s|$ uncorrelated with the system, a unitary
$U^{QE}$, a projector $P^E$ onto the environment alone, and
a constant $c > 0$, such that
\begin{eqnarray} \label{eqtn: general unitary rep}
{\cal E}(\rho) = c \, \mbox{tr}_E (P^E U^{QE} (\rho \otimes \sigma)
	{U^{QE}}^{\dagger} P^E). 
\end{eqnarray}
Furthermore, in the case of a generalized measurement described by
operations ${\cal E}_m$ it is possible to do so in such a way
that for each $m$ the corresponding constant $c_m = 1$, and the projectors
$P^E_m$ form a complete orthogonal set, $\sum_m P^E_m = I$,
$P^E_m P^E_{m'} = \delta_{m,m'} P^E_m$.

Conversely, any map of the form (\ref{eqtn: general unitary rep}) is a quantum
operation.

Once again, introducing a reference system $R$ which purifies $\rho^Q$
we are left with a picture of the dynamics that looks like this:
\begin{figure}
\begin{center}
\unitlength 1cm
\begin{picture}(8,7)(0,0)
\put(2,3){\framebox(1,1){$Q$}}
\put(3,3.5){\vector(1,0){3}}
\put(6,3){\framebox(1,1){$Q'$}}
%
\put(2,5){\framebox(1,1){$R$}}
\put(2.4,4.10){\line(0,1){.05}} \put(2.6,4.10){\line(0,1){.05}}
\put(2.4,4.25){\line(0,1){.05}}  \put(2.6,4.25){\line(0,1){.05}}
\put(2.4,4.40){\line(0,1){.05}}  \put(2.6,4.40){\line(0,1){.05}}
\put(2.4,4.55){\line(0,1){.05}}  \put(2.6,4.55){\line(0,1){.05}}
\put(2.4,4.70){\line(0,1){.05}}  \put(2.6,4.70){\line(0,1){.05}}
\put(2.4,4.85){\line(0,1){.05}}  \put(2.6,4.85){\line(0,1){.05}}
\put(1,4.5){\makebox(0,0){$\left | \Psi^{RQ} \right \rangle$}}
%
\put(4,0.5){\framebox(1,1){$E$}}
\put(4.4,1.7){\vector(0,1){1.6}}
\put(4.6,3.3){\vector(0,-1){1.6}}
\put(4.8,2.5){\makebox(0,0)[cl]{$\sqrt{c} P^E U^{QE}$}}
\end{picture}
\caption{} \label{fig: RQE modified}
\end{center}
\end{figure}

A few miscellaneous remarks will be useful later on.
\begin{enumerate}

\item A prime always denotes a {\em normalized} state. For instance,
\begin{eqnarray}
\rho^{R'Q'} = \frac{({\cal I}_R \otimes {\cal E})(\rho^{RQ})}
	{\mbox{tr}(({\cal I}_R \otimes {\cal E})(\rho^{RQ}))}.
\end{eqnarray}

\item The total state of the system $RQE$ starts and remains pure. That is,
$\rho^{R'Q'E'}$ is a pure state. Purity gives very useful relations 
amongst Von Neumann entropies $S(\rho) \equiv -\mbox{tr}(\rho \log_2 \rho)$,
such as $S(\rho^{R'Q'}) = S(\rho^{E'})$ and all other permutations amongst
$R, Q$ and $E$. These are used frequently in what follows.

\item Generically we denote quantum operations by ${\cal E}$ and the
dimension of the quantum system $Q$ by $d$.

\item {\em Trace-preserving} quantum operations arise when a system interacts
with an environment, and {\em no measurement} is performed on the system
plus environment. Non trace-preserving operations arise when classical
information about the state of the system is made available by
such a measurement. For most of this paper the noisy quantum channel
is described by a trace-preserving quantum operation.

\item Sometimes we consider the composition of two (or more)
quantum operations. Generically we use the notation
${\cal E}_1,{\cal E}_2,\ldots$ for the different operations, and the
notation ${\cal E}_2 \circ {\cal E}_1$ to denote composition of operations,
\begin{eqnarray}
({\cal E}_2 \circ {\cal E}_1)(\rho) \equiv {\cal E}_2 ({\cal E}_1(\rho)).
\end{eqnarray}
Furthermore it is sometimes useful to use the $RQE$ picture of quantum
operations to discuss compositions. We denote the environment
corresponding to operation ${\cal E}_i$ by $E_i$, and assume environments
corresponding to different values of $i$ are independent and initially
pure. So, for example, the initial state for a two-stage composition
would be
\begin{eqnarray}
\rho^{RQE_1E_2} & = & |\psi^{RQ}\rangle \langle \psi^{RQ}| \otimes |s_1\rangle
	\langle s_1| \otimes |s_2 \rangle \langle s_2|. \nonumber \\
 & & \end{eqnarray}
A single prime denotes the state of the system after the application
of ${\cal E}_1$, and a double prime denotes the state of the system
after the application of ${\cal E}_2 \circ {\cal E}_1$, and so on.

\end{enumerate}

\section{Entropy Exchange}
\label{sect: entropy exchange}

This section briefly reviews the definition and some basic
results about the {\em entropy exchange}, which was independently
introduced by Schumacher \cite{Schumacher96a} and Lloyd \cite{Lloyd97a}.
The entropy exchange turns out to be central to understanding the noisy quantum
channel.

The {\em entropy exchange} of a quantum operation ${\cal E}$ with
input $\rho$ is defined to be
\begin{eqnarray}
S_e(\rho,{\cal E}) \equiv S(\rho^{E'}), \end{eqnarray}
where $\rho^{E'}$ is the state of an initially pure environment (the ``mock''
environment of the previous section) after the
operation, and $S(\rho) \equiv -\mbox{tr}(\rho \log_2 \rho)$ is the Von
Neumann entropy. If ${\cal E}(\rho) = \sum_i A_i \rho A_i^{\dagger}$ then a
convenient form for the entropy exchange is found by defining a matrix $W$
with elements
\begin{eqnarray} \label{eqtn: W matrix}
W_{ij} \equiv \frac{\mbox{tr}(A_i \rho A_j^{\dagger})}{\mbox{tr}
	({\cal E}(\rho))}.
\end{eqnarray}
It can be shown \cite{Schumacher96a,Nielsen97a} that
\begin{eqnarray}
S_e(\rho,{\cal E}) = S(W) \equiv - \mbox{tr}(W \log_2 W). \end{eqnarray}
The last equation is frequently useful when performing calculations.

\section{Classical Noisy Channels in a Quantum Setting}
\label{sect: classical}

In this section we show how classical noisy channels can be formulated
in terms of quantum mechanics. We begin by reviewing the formulation in
terms of classical information theory.

A classical noisy channel is described in terms of distinguishable
channel states, which we label by $x$. If the input to the channel is
symbol $x$ then the output is symbol $y$ with probability $p_{y|x}$. 
The channel is assumed to act independently on each input. For
each $x$, the probability sum rule $\sum_y p_{y|x} = 1$ is
satisfied. These {\em conditional probabilities} $p_{y|x}$ completely
describe the classical noisy channel.

Suppose the input to the channel, $x$, is represented by some classical
random variable, $X$, and the output by a random variable $Y$. The
mutual information between $X$ and $Y$ is defined by
\begin{eqnarray}
H(X:Y) \equiv H(X) + H(Y) - H(X,Y),
\end{eqnarray}
where $H(X)$ is the Shannon information of the random variable, $X$,
defined by
\begin{eqnarray}
H(X) \equiv -\sum_x p(x) \log_2 p(x),
\end{eqnarray}
with $0 \log_2 0 \equiv \lim_{p \rightarrow 0}p \log_2 p = 0$.

Shannon showed that the capacity of a noisy classical channel is
given by the expression
\begin{eqnarray}
C_S = \max_{p(x)} H(X:Y), \end{eqnarray}
where the maximum is taken over all possible distributions $p(x)$ for the
channel input, $X$. Notice that although this is not an explicit
expression for the channel capacity in terms of the conditional
probabilities $p_{x|y}$, the maximization can easily be performed
using well known techniques from numerical mathematics. That is,
Shannon's result provides an effective procedure for computing
the capacity of a noisy classical channel.

All these results may be re-expressed in terms of quantum mechanics.
We suppose the channel has some preferred orthonormal basis,
$|x\rangle$, of signal states. For convenience we assume the set of input
states, $|x\rangle$, is the same as the set of output states, $|y\rangle$,
of the channel, although
more general schemes are possible. For the purpose of illustration the
present level of generality suffices. A classical input random variable,
$X$, corresponds to an input density operator for the quantum
channel,
\begin{eqnarray}
\rho_X \equiv \sum_x p(x) |x\rangle \langle x|. \end{eqnarray}
The statistics of $X$ are recoverable by measuring $\rho_X$ in the
$|x\rangle$ basis.
Defining operators $E_{xy}$ by
\begin{eqnarray}
E_{xy} \equiv |y\rangle \langle x|, \end{eqnarray}
we find that the channel operation defined by
\begin{eqnarray}
{\cal N}(\rho) \equiv \sum_{xy} p_{y|x} E_{xy} \rho E_{xy}^{\dagger}.
\end{eqnarray}
is a trace-preserving quantum operation,
and that
\begin{eqnarray}
{\cal N}(\rho_X) = \rho_Y = \sum_y p(y) |y\rangle \langle y|,
\end{eqnarray}
where $\rho_Y$ is the density operator corresponding to the random variable
$Y$ that would have been obtained from $X$ given a classical channel
with probabilities $p_{y|x}$. This gives a quantum mechanical
formalism for describing classical sources and channels. It is interesting to
see what form the mutual information and channel capacity take in the
quantum formalism.

Notice that
\begin{eqnarray}
H(X) & = & S(\rho_X) \\
H(Y) & = & S(\rho_Y) = S({\cal N}(\rho_X)).
\end{eqnarray}
Next we compute the entropy exchange associated with the channel operating
on input $\rho_X$, by computing the $W$ matrix given by (\ref{eqtn: W matrix}).
The $W$ matrix corresponding to the channel with input $\rho_X$ has entries
\begin{eqnarray}
W_{(xy) (x'y')} = \delta_{x,x'} \delta_{y,y'} p(x) p(y|x), \end{eqnarray}
But the joint distribution of $(X,Y)$ satisfies
$p(x) p(y|x) = p(x,y)$. Thus $W$ is diagonal with eigenvalues $p(x,y)$,
so the entropy exchange is given by
\begin{eqnarray}
S_e(\rho_X,{\cal N}) = H(X,Y). \end{eqnarray}
It follows that
\begin{eqnarray}
H(X:Y) = S(\rho_X)+S({\cal N}(\rho_X))-S_e(\rho_X,{\cal N}),
\end{eqnarray}
and thus the Shannon capacity $C_S$ of the classical channel is given in
the quantum formalism by
\begin{eqnarray}
C_S = \max_{\rho_X} \left[ S(\rho_X) + S({\cal N}(\rho_X))-S_e(\rho_X,{\cal N})
\right] ,
\end{eqnarray}
where the maximization is over all input states for the channel, $\rho_X$,
which are diagonal in the $|x\rangle$ basis.

The problem we have been considering is that of transmitting a
discrete set of orthogonal states (the states $|x\rangle$) through the
channel. In many quantum applications one is not only interested in
transmitting a discrete set of states, but also arbitrary superpositions of
those states. That is, one wants to
transmit entire {\em subspaces} of
states. In this case, the capacity of interest is the maximum rate of
transmission of subspace dimensions.
This may occur in quantum computing, cryptography and
teleportation. It is also interesting in these applications to transmit
the {\em entanglement} of states. This cannot be done by considering
the transmission of a set of orthogonal pure states
alone.

It is not difficult to see that $C_S$ is not
correct as a measure of how many subspace dimensions may be reliably
transmitted through a quantum channel. For example consider the classical
noiseless channel,
\begin{eqnarray}
{\cal N}(\rho) = \sum_x |x\rangle \langle x| \rho |x \rangle \langle x|,
\end{eqnarray}
where $|x\rangle$ is an orthonormal set of basis states for the channel.
It is easily seen that
\begin{eqnarray}
C_S = \log_2 d, \end{eqnarray}
where $d$ is the number of channel dimensions. Yet it is intuitively clear,
and is later proved in a more rigorous fashion, that such a
channel cannot be used to transmit any non-trivial subspace of state
space, nor can it be used to transmit any entanglement, and thus its
capacity for transmitting these types of quantum resources is
zero.

\section{Entanglement Fidelity}
\label{sect: entanglement fidelity}

In this section we review a quantity known as the
{\em entanglement fidelity} \cite{Schumacher96a}.
It is this quantity which we
use to study the effectiveness of schemes for sending information
through a noisy quantum channel.

The entanglement fidelity is defined for a {\em process}, specified
by a quantum operation ${\cal E}$ acting on some initial state, $\rho$.
We denote it by $F_e(\rho,{\cal E})$. The concerns motivating the
definition of the entanglement fidelity are twofold:
\begin{enumerate}

\item $F_e(\rho,{\cal E})$ measures how well the {\em state},
$\rho$, is preserved by the operation ${\cal E}$. An entanglement fidelity
close to one indicates that the process preserves the state well.

\item $F_e(\rho,{\cal E})$ measures how well the {\em entanglement} of $\rho$
with other systems is preserved by the operation ${\cal E}$. An
entanglement fidelity close to one indicates the process preserves the
entanglement well.

\end{enumerate}
Conversely, an entanglement fidelity close to zero indicates that 
the state or its entanglement were not well preserved by the operation
${\cal E}$.

Formally, the entanglement fidelity is defined by
\begin{eqnarray}
F_e(\rho,{\cal E}) \equiv \langle \psi^{RQ}| ({\cal I}_R \otimes {\cal E})
	(|\psi^{RQ}\rangle \langle \psi^{RQ}|) |\psi^{RQ} \rangle.
\end{eqnarray}
That is, the entanglement fidelity is the overlap between the initial
purification $|\psi^{RQ}\rangle$ of the state {\em before} it is
sent through the channel with the state of the joint system $RQ$
{\em after} it has been sent through the channel.
The entanglement fidelity depends
only on $\rho$ and ${\cal E}$, not on the particular purification
$|\psi^{RQ}\rangle$ of $\rho$ that is used  \cite{Schumacher96a}. If ${\cal E}$ has operation
elements $\{ A_i \}$ then  
the entanglement fidelity has the expression \cite{Schumacher96a,Nielsen97a}
\begin{eqnarray} \label{eqtn: F_e calculation form}
F_e(\rho,{\cal E}) = \frac{\sum_i | \mbox{tr}(A_i \rho) |^2}
	{\mbox{tr}({\cal E}(\rho))}. \end{eqnarray}
This expression simplifies for trace-preserving quantum operations since
the denominator is one. The entanglement fidelity has the following properties
\cite{Schumacher96a,Schumacher96b,Nielsen97a}.

\begin{enumerate}

\item $0 \leq F_e(\rho,{\cal E}) \leq 1$.

\item $F_e(\rho,{\cal E}) = 1$ if and only if for all pure states
$|\psi\rangle$ lying in the support of $\rho$,
\begin{eqnarray}
{\cal E}(|\psi\rangle \langle \psi|) = |\psi\rangle \langle \psi|.
\end{eqnarray}

\item The entanglement fidelity is a lower bound on the
fidelity defined by Jozsa \cite{Jozsa95a} in the following sense,
\begin{eqnarray}
F_e(\rho,{\cal E}) \leq F(\rho,{\cal E}(\rho)). \end{eqnarray}

\item  Suppose $\{ |\psi_i\rangle, p_i \}$ is an ensemble realizing $\rho$,
\begin{eqnarray}
\rho = \sum_i p_i |\psi_i \rangle \langle \psi_i|. \end{eqnarray}
Then the entanglement fidelity is a lower bound on the
average fidelity for the pure states $|\psi_i\rangle$,
\begin{eqnarray}
F_e(\rho,{\cal E}) \leq \sum_i p_i \langle \psi_i | {\cal E}( |\psi_i\rangle
	\langle \psi_i|) | \psi_i \rangle . \end{eqnarray}

\item Again suppose $\{ |\psi_i\rangle, p_i \}$ is an ensemble realizing $\rho.$
Then if the pure state fidelity
$\langle \psi|{\cal E}(|\psi\rangle \langle \psi |) |\psi\rangle
\ge 1 - \eta$ for all $|\psi\rangle$ in the support of $\rho$,
$F_e(\rho, {\cal E}) \ge 1 - (3/2) \eta.$ 
(Knill and LaFlamme \cite{Knill97a}.)
\end{enumerate}

There are several reasons for using the entanglement fidelity
as our measure of success in transmitting quantum states.  If 
we succeed in sending a source $\rho_s$ with high entanglement
fidelity, we can send {\em any} ensemble for $\rho_s$ with high
average pure-state fidelity, by item 4 above.  Entanglement 
fidelity is thus a more severe requirement of quantum coherence
than average pure-state fidelity.  Moreover, the ability to preserve
entanglement is of great importance in applications of quantum 
coding to, say, quantum computation, where one would like to be 
able to apply error-correction in a modular fashion to small portions
of a quantum computer despite the fact that they may, in the course
of quantum computation, become entangled with other parts
of the computer \cite{Nielsen96c}. (Of course, the general
problem of finding the capacity of a noisy quantum channel for
a {\em given} ensemble with average pure-state fidelity
as the reliability measure is also worth investigating.)  

An appropriate measure of how well a {\em subspace}
of quantum states is transmitted is the {\em subspace fidelity}
\begin{eqnarray} \label{eqtn: subspace fidelity}
F_s(P,{\cal E}) \equiv \min_{|\psi\rangle} \langle \psi | 
{\cal E}(|\psi \rangle \langle \psi|)|\psi \rangle,
\end{eqnarray}
where the minimization is over all pure states $|\psi\rangle$ in 
the subspace whose projector is $P.$  Item 5 above implies that
if the subspace fidelity is close to one, the entanglement fidelity
is also close to one. The converse is not in general true.
That is, reliable transmission of subspaces is a more stringent
requirement than transmission of entanglement.
Therefore using entanglement fidelity as our criterion for
reliable transmission yields capacities at least as great as those obtained
when subspace fidelity is used as the criterion.  We conjecture
that these two capacities are identical.

As an alternative measure of subspace fidelity, one might consider
the average pure state fidelity 
\begin{eqnarray}
\int d|\psi\rangle \langle \psi |
{\cal E}(|\psi \rangle \langle \psi |)|\psi \rangle,
\end{eqnarray}
where the integration is done using the unitarily invariant
measure on the subspace of interest.  By item 4 above, the capacity
resulting from this measure of reliability is at least as great
as that which results when entanglement fidelity is used as the
measure of reliability.  We do not know whether these two capacities
are equal.

The lesson to be learnt from this discussion is that there are
many different measures which may be used to quantify how reliably
quantum states are transmitted, and different measures may result
in different capacities.  Which measure is used depends on what
resource is most important for the application of interest.  For
the remainder of this paper, we use the entanglement fidelity
as our measure of reliability.

There is a very useful inequality, the {\em quantum Fano inequality},
which relates the entropy exchange and the entanglement fidelity. It
is \cite{Schumacher96a}:
\begin{eqnarray} \label{eqtn: quantum Fano}
S_e(\rho,{\cal E}) & \leq & h(F_e(\rho,{\cal E})) + (1-F_e(\rho,{\cal E}))
  \log_2 (d^2-1), \nonumber \\
& &
\end{eqnarray}
where $h(p) \equiv -p \log p - (1-p) \log (1-p)$ is the dyadic Shannon
information associated with $p$.
It is useful to note for our later work that
$0 \leq h(p) \leq 1$ and $\log (d^2-1) \leq 2 \log d$, so from the
quantum Fano inequality,
\begin{eqnarray} \label{eqtn: quantum Fano 2}
S_e(\rho,{\cal E}) & \leq & 1 + 2(1-F_e(\rho,{\cal E}))
  \log d. \nonumber \\
& &
\end{eqnarray}

The proof of the quantum Fano inequality, (\ref{eqtn: quantum Fano}), is
simple enough that for convenience we repeat it here. Consider
an orthonormal set of $d^2$ basis states, $|\psi_i\rangle$, for the system
$RQ$. This basis set is chosen so \linebreak $|\psi_1\rangle = |\psi^{RQ}\rangle$. If we
form the quantities $p_i \equiv \langle \psi_i| \rho^{R'Q'} | \psi_i \rangle$,
then it is possible to show (see for example \cite{Wehrl78a}, page 240)
\begin{eqnarray}
S(\rho^{R'Q'}) \leq H(p_1,\ldots,p_{d^2}), \end{eqnarray}
where $H(p_i)$ is the Shannon information of the set $p_i$. But by easily
verified grouping properties of the Shannon entropy,
\begin{eqnarray}
\label{eqtn: entropy grouping}
H(p_1,\ldots,p_{d^2})  & = &  h(p_1) +
 (1-p_1)H(\frac{p_2}{1-p_1},\ldots,\frac{p_{d^2}}{1-p_1}), \nonumber \\
 & & \end{eqnarray}
and it is easy to show that $H(\frac{p_2}{1-p_1},\ldots,\frac{p_{d^2}}{1-p_1}) \leq \log (d^2-1)$. Combining these
results and noting that $p_1 = F_e(\rho,{\cal E})$ by definition of the
entanglement fidelity,
\begin{eqnarray}
S_e(\rho,{\cal E}) & \leq & h(F_e(\rho,{\cal E})) + (1-F_e(\rho,{\cal E}))
  \log (d^2-1), \nonumber \\
& & 
\end{eqnarray}
which is the quantum Fano inequality.

For applications it is useful to understand the continuity properties of the
entanglement fidelity. To that end we prove the following lemma:

{\em Lemma} (continuity lemma for entanglement fidelity)

Suppose ${\cal E}$ is a trace-preserving quantum operation, $\rho$ is
a density operator, and $\Delta$ is a Hermitian operator with trace zero. Then
\begin{eqnarray} \label{eqtn: continuity lemma}
|F_e(\rho+\Delta,{\cal E}) - F_e(\rho,{\cal E})| & \leq &   2 \mbox{tr}(|\Delta |)
	+ \mbox{tr}(|\Delta |)^2. \nonumber \\
& & 
\end{eqnarray}

To prove the lemma we apply (\ref{eqtn: F_e calculation form}) to obtain
\begin{eqnarray}
|F_e(\rho+\Delta,{\cal E})-F_e(\rho,{\cal E})| & \leq & 
	 2 \sum_i |\mbox{tr}(A_i \rho)|\,| \mbox{tr}(A_i^{\dagger} \Delta)| + 
\nonumber \\
& &	\sum_i |\mbox{tr}(A_i \Delta)|^2 . 
\end{eqnarray}
Applying a Cauchy-Schwarz inequality to each sum, the first with respect
to the complex inner product $\sum_i x_i^* y_i$, the second with respect
to the Hilbert-Schmidt inner product $\mbox{tr}(X^{\dagger} Y)$,
we obtain
\begin{eqnarray}
|F_e(\rho+\Delta,{\cal E})-F_e(\rho,{\cal E})| \leq & & \nonumber \\
	 2 \left( \sum_i |\mbox{tr}(A_i \rho)|^2 \sum_j
	|\mbox{tr}(A_j^{\dagger} \Delta)|^2 \right)^{\frac{1}{2}} + & & \nonumber \\
	\sum_i |\mbox{tr}(A_i | \Delta | A_i^{\dagger})| \,
	| \mbox{tr}( |\Delta |) |, & & \end{eqnarray}
where $|\Delta | \equiv \sqrt{\Delta^{\dagger} \Delta}$. 
Applying (\ref{eqtn: F_e calculation form}) and $F_e(\rho,{\cal E}) \leq 1$
to the first sum and the trace-preserving property of ${\cal E}$ to
the final sum gives
\begin{eqnarray}
|F_e(\rho+\Delta,{\cal E})-F_e(\rho,{\cal E})| & \leq &
	 2 \sqrt{\sum_j |\mbox{tr}(A_j^{\dagger} \Delta)|^2 } +
	\mbox{tr}(|\Delta |)^2. \nonumber \\
& & \end{eqnarray}
One final application of the Cauchy-Schwarz inequality and the
trace-preserving property of ${\cal E}$ gives
\begin{eqnarray}
|F_e(\rho+\Delta,{\cal E})-F_e(\rho,{\cal E})| & \leq &
	 2 \mbox{tr}(|\Delta |) + \mbox{tr}(|\Delta |)^2, \end{eqnarray}
as required.

This result gives bounds on the change in the entanglement fidelity when
the input state is perturbed. Note, incidentally, that during the proof
a coefficient $\sqrt{F_e(\rho,{\cal E})}$ was dropped from the first term on
the right hand side of the inequality. For some applications it may be
useful to apply the inequality with this coefficient in place.

\section{Coherent Information}
\label{sect: coherent information}

In this section we investigate the {\em coherent information}. 
The coherent
information was defined in \cite{Schumacher96b}, where it was 
suggested that
the coherent information plays a role in quantum information 
theory analogous
to the role played by mutual information in classical information theory
in the following sense. Consider a classical random process,
\begin{eqnarray} \label{eqtn: classical process}
X \stackrel{{\cal M}}{\rightarrow} Y, \end{eqnarray}
in which the random variable $X$ is used as the input to some 
process which produces as output the random variable $Y$. The
distributions of $X$ and $Y$ are related by a linear map, ${\cal M}$,
determined by the conditional probabilities of the process. 
An example of such a process is a noisy classical channel with input $X$
and output $Y$. As discussed earlier, an important
quantity in information theory is the mutual information, $H(X:Y)$,
between the input $X$ and the output $Y$ of the process. Note that
$H(X:Y)$ can be regarded as a function of the input $X$ and
the map ${\cal M}$ only, since the joint distribution of $X$ and $Y$
is determined by these.

Quantum mechanically we can consider
a process defined by an input $\rho$, and output $\rho'$, with the process
described by a quantum operation, ${\cal E}$,
\begin{eqnarray} \label{eqtn: quantum process}
\rho \stackrel{{\cal E}}{\rightarrow} \rho' = {\cal E}(\rho). \end{eqnarray}
We assert that the coherent information, defined by
\begin{eqnarray}
I(\rho,{\cal E}) \equiv S \left(
	\frac{{\cal E}(\rho)}{\mbox{tr}({\cal E}(\rho))} \right) -
	S_e(\rho,{\cal E}),
\end{eqnarray}
plays a role in quantum information theory analogous to that played
by the mutual information $H(X:Y)$ in classical information theory. This
is not obvious from the definition, and one goal of this section
is to make it appear plausible that this is the case. Of course, the true
justification for regarding the coherent information as the quantum
analogue of the mutual information is its success as the quantity
appearing in results on channel capacity, as discussed in later
sections. This is the true motivation for all definitions in information
theory, whether classical or quantum: their success at quantifying
the  resources needed to perform some interesting physical task,
not some abstract mathematical motivation.

In subsection \ref{subsect: data processing inequality} we review
the data processing inequality which provides motivation
for regarding the coherent information as a quantum analogue of the
mutual information, and whose application is crucial to later
reasoning. Subsection \ref{subsect: I properties} studies in detail
the properties of the coherent information. In particular, we
prove several results related to convexity that are useful both as
calculational aids, and also for proving later results. 
Subsection \ref{subsect: fidelity lemma} proves a lemma
about the entanglement fidelity that glues together many of our later proofs
of upper bounds on the channel capacity. Finally, subsections
\ref{subsect: example 1} and \ref{subsect: example 2} describe
two important ways the behaviour of the coherent information differs
from the behaviour of the mutual information when quantum entanglement
is allowed.

\subsection{Quantum Data Processing Inequality}
\label{subsect: data processing inequality}

The role of coherent information in quantum information theory is intended
to be similar to that of mutual information  in classical information
theory. This is not obvious from the definition, but can be given
an operational motivation in terms of a procedure known as
{\em data processing}. The classical data processing inequality
\cite{Cover91a} states that any three variable Markov process,
\begin{eqnarray}
X \rightarrow Y \rightarrow Z, \end{eqnarray}
satisfies a data processing inequality,
\begin{eqnarray}
H(X) \geq H(X:Y) \geq H(X:Z). \end{eqnarray}
The idea is that the operation $Y \rightarrow Z$ represents 
some kind of ``data processing'' of $Y$ to obtain $Z$, and 
the mutual information after processing, $H(X:Z)$, can be no higher
than the mutual information before processing, $H(X:Y)$. Furthermore,
suppose we have a Markov process,
\begin{eqnarray}
X \rightarrow Y, \end{eqnarray}
such that $H(X) = H(X:Y)$. Intuitively, one might expect that it be
possible to do data processing on $Y$ to recover $X$.
It is not difficult to show that it is possible, using $Y$ alone,
to construct a third variable, $Z$, forming a third stage in
the Markov process,
\begin{eqnarray}
X \rightarrow Y \rightarrow Z, \end{eqnarray}
such that $X = Z$ with probability one, if and only if $H(X) = H(X:Y)$.

An analogous quantum result has been proved by Schumacher and Nielsen
\cite{Schumacher96b}. It states that given trace-preserving
quantum operations ${\cal E}_1$ and ${\cal E}_2$ defining a quantum
process,
\begin{eqnarray} 
\rho \rightarrow {\cal E}_1(\rho) \rightarrow
	 ({\cal E}_2 \circ {\cal E}_1)(\rho), \end{eqnarray}
then
\begin{eqnarray} \label{eqtn: data processing}
S(\rho) \geq I(\rho,{\cal E}_1) \geq I(\rho,{\cal E}_2 \circ {\cal E}_1).
\end{eqnarray}
Furthermore, it was shown in \cite{Schumacher96b} that given a process
\begin{eqnarray}
\rho \rightarrow {\cal E}_1(\rho), \end{eqnarray}
it is possible to find an operation ${\cal E}_2$ which reverses
${\cal E}_1$ if and only if
\begin{eqnarray}
S(\rho) = I(\rho,{\cal E}_1). \end{eqnarray}
The close analogy between the classical and quantum data 
processing inequalities
provides a strong operational motivation for considering
the coherent information to be the quantum analogue of the
classical mutual information.

The proof of the quantum data processing inequality
is repeated here in order
to address the issue of what happens when ${\cal E}_1$ and ${\cal E}_2$
are not trace-preserving.
The proof of the first inequality is to apply the subadditivity inequality
\cite{Wehrl78a} $S(\rho^{R'E'}) \leq S(\rho^{R'}) + S(\rho^{E'})$ in
the $RQE$ picture of operations to obtain
\begin{eqnarray}
I(\rho,{\cal E}_1) & = & S({\cal E}_1(\rho)) - S_e(\rho,{\cal E}_1) \\
 & = & S(\rho^{Q'}) - S(\rho^{E'}) \\
 & = & S(\rho^{R'E'})-S(\rho^{E'}) \\
 & \leq & S(\rho^{R'}) = S(\rho^R) = S(\rho).
\end{eqnarray}
It is clear that this part of the inequality need not hold
when ${\cal E}_1$
is not trace-preserving. The reason for this is that it is no
longer necessarily the case that $\rho^{R'} = \rho^R$, and thus
it may not be possible to make the identification 
$S(\rho^{R'}) = S(\rho^R)$. For example, suppose 
we have a three dimensional state space
with orthonormal states $|1\rangle$, $|2\rangle$ and $|3\rangle$. 
Let $P_{12}$
be the projector onto the two dimensional subspace spanned by
$|1\rangle$ and $|2\rangle$, and $P_3$ the projector onto the
subspace spanned by $|3\rangle$. Let $\rho = \frac{p}{2} P_{12} +
(1-p)P_3$, where $0 < p < 1$, and
${\cal E}(\rho) = P_{12} \rho P_{12}$. Then by choosing $p$
small enough we can make $S(\rho) \approx 0$, but
$I(\rho,{\cal E}) = 1$, so we have an example of a
non trace-preserving operation which does not obey the data
processing inequality.

The proof of the second part of the data processing inequality is to apply
the strong subadditivity inequality \cite{Wehrl78a},
\begin{eqnarray} \label{eqtn: strong subadditivity}
S(\rho^{R''E_1''E_2''}) + S(\rho^{E_1''}) \leq S(\rho^{R''E_1''})
	+ S(\rho^{E_1'' E_2''}),
\end{eqnarray}
where we are now using an $RQE_1E_2$ picture of the operations.
{}From purity of the total state of $RQE_1E_2$ it follows that
\begin{eqnarray}
S(\rho^{R''E_1''E_2''}) = S(\rho^{Q''}). \end{eqnarray}
Neither of the systems $R$ or $E_1$ are involved in the second stage of
the dynamics in which $Q$ and $E_2$ interact unitarily. Thus, their state
does not change during this stage: $\rho^{R''E_1''} = \rho^{R'E_1'}$.
But from the purity of $RQE_1$ after the first stage of the dynamics,
\begin{eqnarray}
S(\rho^{R''E_1''}) = S(\rho^{R'E_1'}) = S(\rho^{Q'}).
\end{eqnarray}
The remaining two terms in the subadditivity inequality are now recognized
as entropy exchanges,
\begin{eqnarray} 
S(\rho^{E_1''}) = S(\rho^{E_1'}) = S_e(\rho,{\cal E}_1), \\
S(\rho^{E_1''E_2''}) = S_e(\rho,{\cal E}_2 \circ {\cal E}_1). 
\end{eqnarray}
Making these substitutions into the inequality obtained from
strong subadditivity (\ref{eqtn: strong subadditivity})
yields
\begin{eqnarray}
S(\rho^{Q''}) + S_e(\rho,{\cal E}_1) \leq S(\rho^{Q'}) + S_e(\rho,
	{\cal E}_2 \circ {\cal E}_1), 
\end{eqnarray}
which can be rewritten as the second stage of the data processing
inequality,
\begin{eqnarray}
I(\rho,{\cal E}_1) \geq I(\rho,{\cal E}_2 \circ {\cal E}_1).
\end{eqnarray}

Notice that this inequality holds provided ${\cal E}_2$ is trace-preserving,
and does not require any assumption that ${\cal E}_1$ is trace-preserving.
This is very useful in our later work.

\subsection{Properties of Coherent Information}
\label{subsect: I properties}

The set of completely positive maps forms a positive cone, that is, if ${\cal E}_i$
is a collection of completely positive maps and $\lambda_i$ is a set
of non-negative numbers then $\sum_i \lambda_i {\cal E}_i$ is
also a completely positive map. In this
section we prove two very useful properties of the coherent information. First,
it is easy to see that for any quantum operation ${\cal E}$ and non-negative
$\lambda$,
\begin{eqnarray}
I(\rho,\lambda {\cal E}) = I(\rho,{\cal E}). \end{eqnarray}
This follows immediately from the definition of the coherent information.
A slightly more difficult property to prove is the following.

\newpage

{\em Theorem} (generalized convexity theorem for coherent information).

Suppose ${\cal E}_i$ are quantum operations. Then
\begin{eqnarray} \label{eqtn: abstract subadditivity}
I(\rho,\sum_i {\cal E}_i) \leq \frac{\sum_i \mbox{tr}({\cal E}_i(\rho))
	I(\rho,{\cal E}_i)}{\mbox{tr}(\sum_i {\cal E}_i(\rho))}.
\end{eqnarray}

This result is extremely useful in  our later work. An important
and immediate corollary is the following:

{\em Corollary} (convexity theorem for coherent information)

If a trace-preserving operation,
${\cal E} = \sum_i p_i {\cal E}_i$ is a convex sum
($p_i \geq 0, \sum_i p_i = 1$) of trace-preserving
operations ${\cal E}_i$, then the coherent information is
convex,
\begin{eqnarray}
I(\rho,\sum_i p_i {\cal E}_i) \leq \sum_i p_i I(\rho,{\cal E}_i).
\end{eqnarray}

The proof of the corollary is immediate from the theorem.
The theorem follows from the concavity of the {\em conditional}
entropy (see references cited
in \cite{Wehrl78a}, pages 249--250), which for two
systems, $1$ and $2$, is defined by
\begin{eqnarray}
S(2|1) \equiv S(\rho_{12})-S(\mbox{tr}_2(\rho_{12})). \end{eqnarray}
This expression is concave in $\rho_{12}$. Now notice that
\begin{eqnarray}
I(\rho,{\cal E}) = S(\rho^{Q'})-S(\rho^{R'Q'}) = -S(R' | Q').
\end{eqnarray}
The theorem now follows from the concavity of the conditional
entropy.





A further useful result concerns the additivity of coherent
information,

{\em Theorem} (additivity for independent channels) 

Suppose ${\cal E}_1,\ldots,{\cal E}_n$ are quantum operations and
$\rho_1,\ldots,\rho_n$ are density operators. Then
\begin{eqnarray}
I(\rho_1 \otimes \ldots \otimes \rho_n,{\cal E}_1 \otimes \ldots {\cal E}_n)
	= \sum_i I(\rho_i,{\cal E}_i). \end{eqnarray}
The proof is immediate from the additivity property of entropies
for product states.

\subsection{A Lemma About Entanglement Fidelity}
\label{subsect: fidelity lemma}

The following lemma is the glue which holds together much of our later
work on proving upper bounds to channel capacities. In this section
we prove the lemma only for the special case of trace-preserving
operations. A similar but more complicated result is true for general
operations, and is given in section \ref{sect: observed channel}.

We begin by repeating the proof of a simple inequality that was first
proved in \cite{Schumacher96a}, which states that the decrease (if any)
in system entropy must be bounded above by the increase in the entropy
of a pure environment. This applies only for trace-preserving
operations ${\cal E}$. Applying the subadditivity inequality
\cite{Wehrl78a} $S(\rho^{Q'E'}) \leq S(\rho^{Q'}) + S(\rho^{E'})$ and
the relationship $S(\rho^{R'}) = S(\rho^{Q'E'})$ which follows from
purity we obtain
\begin{eqnarray}
S(\rho) & = & S(\rho^R) \\
  & = & S(\rho^{R'}) \\
 & = & S(\rho^{Q'E'}) \\
 & \leq & S(\rho^{Q'}) + S(\rho^{E'}).
\end{eqnarray}
Rewriting this slightly gives
\begin{eqnarray} \label{eqtn: Araki-Lieb}
S(\rho)-S({\cal E}(\rho)) \leq S_e(\rho,{\cal E}), \end{eqnarray}
for any trace-preserving quantum operation ${\cal E}$.

{\em Lemma} (entanglement fidelity lemma for operations)

Suppose ${\cal E}$ is
a trace-preserving quantum operation, and $\rho$ is some quantum
state. Then for all trace-preserving quantum operations ${\cal D}$,
\begin{eqnarray} \label{eqtn: fidelity lemma}
S(\rho) \leq I(\rho,{\cal E}) + 2
	+ 4 (1-F_e(\rho,{\cal D} \circ {\cal E})) \log d.
\end{eqnarray}

This lemma is extremely useful in obtaining proofs of bounds on the channel
capacity. In order for the entanglement fidelity
to be close to one, the quantity appearing on the right hand side must
be close to zero. This shows that the entropy of $\rho$ cannot
greatly exceed the coherent information $I(\rho,{\cal E})$ if the
entanglement fidelity is to be close to one.

To prove the lemma, notice that by the second part of
the data processing inequality, (\ref{eqtn: data processing}),
\begin{eqnarray}
S(\rho)-I(\rho,{\cal E}) & \leq & S(\rho) - S(({\cal D} \circ
	{\cal E})(\rho)) +
	S_e(\rho,{\cal D} \circ {\cal E}). \nonumber \\
& & \end{eqnarray}
Applying inequality (\ref{eqtn: Araki-Lieb}) gives
\begin{eqnarray}
S(\rho)-S(({\cal D} \circ {\cal E})(\rho))
    \leq S_e(\rho,{\cal D} \circ {\cal E}), \end{eqnarray}
and combining the last two inequalities gives
\begin{eqnarray}
S(\rho)-I(\rho,{\cal E}) & \leq & 2 S_e(\rho,{\cal D} \circ {\cal E}) \\
 & \leq & \label{eqtn: stronger bounds}
	2 h(F_e(\rho,{\cal D} \circ {\cal E})) + \nonumber \\
 & & 2(1-F_e(\rho,{\cal D} \circ {\cal E})) \log (d^2-1),
\end{eqnarray}
where the second step follows from the quantum Fano inequality
(\ref{eqtn: quantum Fano}). But the dyadic Shannon entropy $h$ is bounded
above by $1$ and $\log (d^2-1) \leq 2 \log d$, so
\begin{eqnarray}
S(\rho) \leq I(\rho,{\cal E})+2+4 (1-F_e(\rho,{\cal D} \circ {\cal E}))
	\log d. \end{eqnarray}
This completes the proof.

This inequality is strong enough to prove the asymptotic bounds
which are of most interest for our later work. The somewhat stronger
inequality (\ref{eqtn: stronger bounds}) is also useful when proving
one-shot results, that is, when no block coding is being used.

\subsection{Quantum Characteristics of the Coherent Information I}
\label{subsect: example 1}

There are at least two important respects 
in which the coherent information behaves
differently from the classical mutual information. In this subsection and
the next we explain what these differences are.

Classically, suppose we have a Markov process,
\begin{eqnarray}
X \rightarrow Y \rightarrow Z. \end{eqnarray}
Intuitively we expect that
\begin{eqnarray}
H(X:Z) \leq H(Y:Z), \end{eqnarray}
and, indeed, it is not difficult
to prove such a ``pipelining inequality'',
based on the definition of the mutual information. The idea is that
any information about $X$ that reaches $Z$ must go through $Y,$ and therefore
is 
also information that $Z$ has about $Y$.
However, the quantum mechanical analogue of this result fails to hold.
We shall see that the reason it fails is due to quantum
entanglement.

{\bf Example 1:}

Suppose we have a two-part quantum process described by quantum 
operations
${\cal E}_1$ and ${\cal E}_2$.
\begin{eqnarray}
\rho \rightarrow {\cal E}_1 (\rho) \rightarrow 
({\cal E}_2 \circ {\cal E}_1)
	(\rho). \end{eqnarray}
Then, in general
\begin{eqnarray} \label{eqtn: example 1}
I(\rho,{\cal E}_2 \circ {\cal E}_1) \not\leq 
I({\cal E}_1(\rho),{\cal E}_2).
\end{eqnarray}
An explicit example showing that this is the case is given below.  
It is not possible to prove a general inequality of this
sort for the coherent information - examples may be found where a 
$<,>$ or $=$ sign could occur in the last equation.
We now show how the purely quantum mechanical effect of
entanglement is responsible for this property
of coherent information.

Notice that the truth of the equation
\begin{eqnarray} \label{eqtn: qm subadditivity}
I(\rho,{\cal E}_2 \circ {\cal E}_1) \leq
I({\cal E}_1(\rho),{\cal E}_2),
\end{eqnarray}
is equivalent to 
\begin{eqnarray}
S_e({\cal E}_1(\rho),{\cal E}_2) \leq S_e(\rho,{\cal E}_2 \circ {\cal E}_1).
\end{eqnarray}
This last equation makes it easy to see why 
(\ref{eqtn: qm subadditivity}) may fail. It is because the entropy of the
joint environment for processes ${\cal E}_1$ and ${\cal E}_2$ (the
quantity on the right-hand side) may be less than the entropy of
the environment for process ${\cal E}_2$ alone (the quantity on the left).
This is a property peculiar to quantum mechanics, which is caused
by entanglement; there is no classical analogue. An example of this
type of phenomenon is provided by an EPR pair, 
where the entropy of either
system alone (one bit) is greater than that 
of the entire system, which is
pure and thus has zero bits of entropy.

An example of (\ref{eqtn: example 1}) is as follows. For convenience
we use the language of coding and channel 
operations, since that
language is most convenient later. ${\cal E}_1$ is
to be identified with the coding operation, ${\cal C}$, and
${\cal E}_2$ is to be identified with 
the channel operation, ${\cal N}$.

Suppose we have
a four dimensional state space. We suppose that we have an orthonormal
basis $|1\rangle,|2\rangle,|3\rangle,|4\rangle$, 
and that $P_{12}$ is the
projector onto the space spanned by $|1\rangle$ and $|2\rangle$, and
$P_{34}$ is the projector onto the space spanned by $|3\rangle$ and
$|4\rangle$. Let $U$ be a unitary operator defined by
\begin{eqnarray}
U \equiv |3\rangle \langle 1 | + |4\rangle \langle 2| + 
|1\rangle \langle 3|
	+ |2\rangle \langle 4| . \end{eqnarray}
The channel operation is
\begin{eqnarray}
{\cal N}(\rho) = P_{12} \rho P_{12} + U^{\dagger} 
P_{34} \rho P_{34} U,
\end{eqnarray}
and we use an encoding defined by
\begin{eqnarray}
{\cal C}(\rho) = \frac{1}{2} P_{12}\rho P_{12} + 
\frac{1}{2} U P_{12} \rho
	P_{12} U^{\dagger} + P_{34} \rho P_{34}.
\end{eqnarray}
It is easily checked that for any state $\rho$ whose support lies
wholly in the space spanned by $|1\rangle$ and $|2\rangle$,
\begin{eqnarray}
({\cal N} \circ {\cal C})(\rho) = \rho.
\end{eqnarray}
It follows that
\begin{eqnarray}
I(\rho,{\cal N} \circ {\cal C}) = S(\rho).
\end{eqnarray}
It is also easy to verify that
\begin{eqnarray}
I({\cal C}(\rho),{\cal N}) = 2 S(\rho) - 1.
\end{eqnarray}
Thus there exist states $\rho$ such that
\begin{eqnarray}
I(\rho,{\cal N} \circ {\cal C}) > I({\cal C}(\rho),{\cal N}), \end{eqnarray}
providing an example of (\ref{eqtn: example 1}).

\subsection{Quantum Characteristics of the Coherent Information II}
\label{subsect: example 2}

\begin{figure}
\epsfxsize 3.4in
\epsfbox{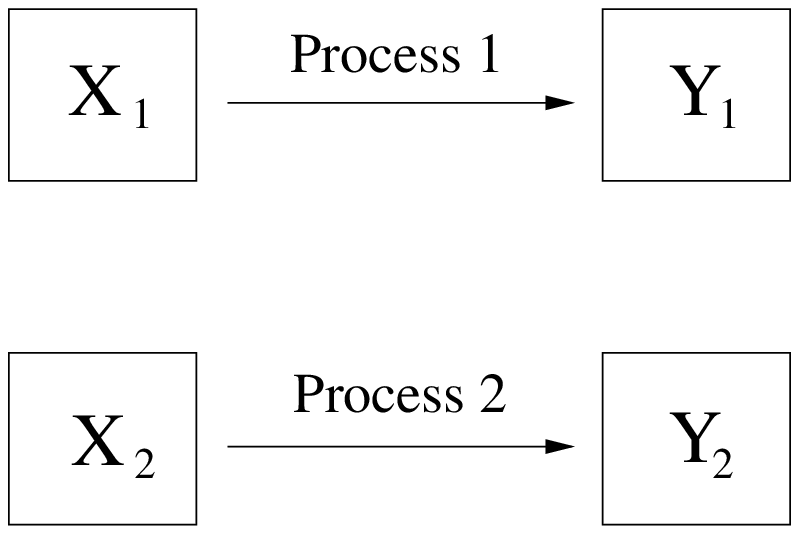}
\caption{} \label{fig: subadditivity}
Dual classical channels operating on inputs $X_1$ and $X_2$
produce outputs $Y_1$ and $Y_2$.
\end{figure}

The second important difference between coherent information and
classical mutual information is related to the property known classically
as {\em subadditivity of mutual information}. Suppose we have several
independent channels operating. Figure \ref{fig: subadditivity} shows
the case of two channels.

These channels are numbered $1,\ldots,n$ and take as inputs random
variables $X_1,\ldots,X_n$. The channels might be separated spatially,
as shown in the figure, or in time. The channels are assumed to act
independently on their respective inputs, and produce
outputs $Y_1,\ldots,Y_n$. It is not difficult
to show that
\begin{eqnarray} \label{eqtn: classical subadditivity}
H(X_1,\ldots,X_n : Y_1,\ldots,Y_n) \leq \sum_i H(X_i : Y_i).
\end{eqnarray}
This property is known as the {\em subadditivity} of mutual information.
It is used, for example, in proofs of the weak converse to Shannon's
noisy channel coding theorem. We now show that the corresponding
quantum statement about coherent information fails to hold.

{\bf Example 2:}
There exists a quantum operation ${\cal E}$ and
a density operator $\rho_{12}$ such that
\begin{eqnarray} \label{eqtn: example 2}
I(\rho_{12},{\cal E} \otimes {\cal E}) \not \leq 
	I(\rho_1,{\cal E}) + I(\rho_2,{\cal E}), \end{eqnarray}
where $\rho_1 \equiv \mbox{tr}_2(\rho_{12})$ and
$\rho_2 \equiv \mbox{tr}_1(\rho_{12})$ are the usual 
reduced density operators
for systems $1$ and $2$.

An example of (\ref{eqtn: example 2}) is 
the following. Suppose system
$1$ consists of two qubits, $A$ and $B$. System 
$2$ consists of two more
qubits, $C$ and $D$. As the initial state we choose
\begin{eqnarray}
\rho_{12} = \frac{I_A}{2} \otimes |\psi^{BD}\rangle 
\langle \psi^{BD}| \otimes
	\frac{I_C}{2}, \end{eqnarray}
where $|\psi^{BD}\rangle$ is a Bell state shared 
between systems $B$ and
$D$.

The action of the channel on $A$ and $B$ is as follows: it sets bit $B$
to some standard state, $|0\rangle$, and allows $A$ through unchanged.
This is achieved by swapping the state of $B$ out into the environment.
Formally,
\begin{eqnarray}
{\cal E}(\rho_{AB}) = \rho_A \otimes |0 \rangle \langle 0 |.
\end{eqnarray}
The same channel is now set to act on systems $C$ and $D$:
\begin{eqnarray}
{\cal E}(\rho_{CD}) = \rho_C \otimes |0 \rangle \langle 0 |.
\end{eqnarray}
A straightforward though slightly tedious 
calculation shows that with this
channel setup
\begin{eqnarray}
I(\rho_1,{\cal E}) = I(\rho_2,{\cal E}) = 0, \end{eqnarray}
and
\begin{eqnarray}
I(\rho_{12},{\cal E} \otimes {\cal E}) = 2. \end{eqnarray}
Thus this setup provides an example of (\ref{eqtn: example 2}).

\section{Noisy Channel Coding Revisited}
\label{sect: noisy coding revisited}

\begin{figure}
\epsfxsize 3.4in
\epsfbox{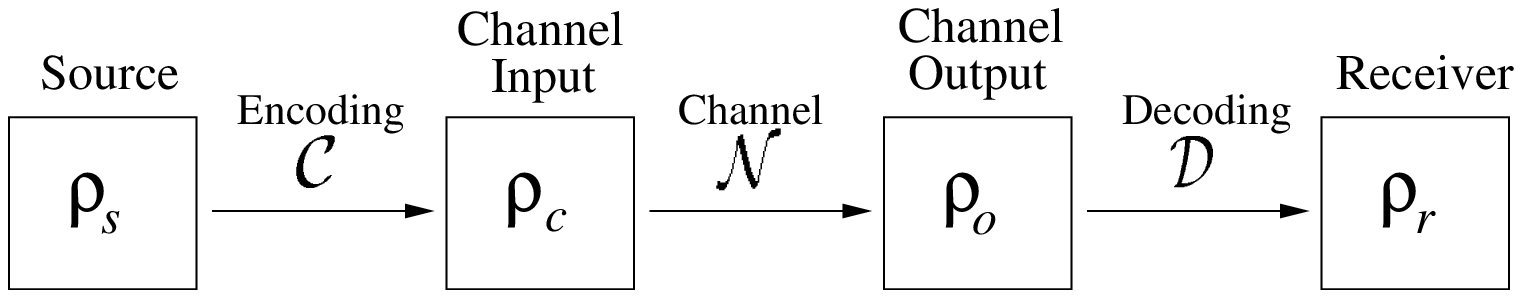}
\caption{} \label{fig: channel1}
The noisy quantum channel, together with encodings and decodings.
\end{figure}

In this section we return to noisy 
channel coding. Recall the basic
procedure for noisy channel coding, as 
illustrated in figure \ref{fig: channel1}.

Suppose a quantum source has output
$\rho_s$. A quantum operation, which we shall denote ${\cal C}$, is
used to {\em encode} the source source, giving the input state to the
channel, $\rho_c \equiv {\cal C}(\rho_s)$.
The encoded state is used as input
to the noisy channel, giving a channel output $\rho_o \equiv
{\cal N}(\rho_c)$. Finally, a decoding quantum operation, ${\cal D}$, is
used to decode the
output of the channel, giving a {\em received state},
$\rho_r \equiv {\cal D}(\rho_o)$. The goal of noisy channel
coding is to find out what source states can be sent with high entanglement
fidelity. That is, we want to know for what states $\rho_s$ can encoding and
decoding operations be found such that
\begin{eqnarray}
F_e(\rho_s,{\cal D} \circ {\cal N} \circ {\cal C}) \approx 1. 
\end{eqnarray}
If large blocks of source states with entropy 
$R$ per use of the channel can be sent through the channel 
with high fidelity, we say the
channel is transmitting at the rate $R.$  

Shannon's noisy channel coding theorem is an example of a
{\em channel capacity} theorem. Such theorems come in two parts:
\begin{enumerate}

\item An {\em upper bound} is placed on the rate at which information
can be sent reliably through the channel. There should be an effective
procedure for calculating this upper bound.

\item It is proved that a reliable
scheme for encoding and decoding exists which
comes arbitrarily close to {\em attaining} the upper bound found in 1.

\end{enumerate}
This maximum rate at which information can be reliably sent through the
channel is known as the {\em channel capacity}.

In this paper we consider only the first of these tasks, the
placing of upper bounds on the rate at which quantum information
can be reliably sent through a noisy quantum channel.
The results we prove are analogous to the
weak converse of the classical noisy coding theorem, but cannot be
considered true converses until attainability of our bounds is demonstrated.
We do consider it likely that our bounds are equal to the true quantum 
channel capacity.


\subsection{Mathematical Formulation of Noisy Channel Coding}
\label{subsect: other stuff}

Up to this point the procedure for doing noisy channel coding has been
discussed in broad outline but we have not made all of our definitions
mathematically precise. This subsection gives a precise
formulation for the most important concepts appearing in our work on
noisy channel coding.

Define a {\em quantum source} $\Sigma = (H_s,\Upsilon)$ to consist of a
Hilbert space $H_s$ and a sequence $\Upsilon =
\{\rho_s^1,\rho_s^2,...,\rho_s^n,...\}$ where $\rho_s^1$ is a density 
operator on $H_s$, $\rho_s^2$ a density operator on $H_s \otimes
H_s$, and $\rho_s^n$ a density operator on $H_s^{\otimes n},$ etc... 
Using, for example,  ``${\rm tr}_{34}$'' to denote the partial trace over the third
and fourth copies of $H_s,$ we require as part of our definition
of a quantum source that for all $j$ and all $n > j,$
\begin{eqnarray}
{\rm tr}_{j+1,...,n} (\rho_s^n) = \rho_s^j,
\end{eqnarray}
i.e. that density operators in the sequence be consistent with
each other in the sense that earlier ones be derivable from later
ones by an appropriate partial trace.  The \linebreak 
$n$-th density operator
is meant to represent the state of $n$ emissions from the source,
normally thought of as taking $n$ units of time.
(We could have used a single density operator on a countably infinite
tensor product of spaces $H_s,$ but we wish to avoid the technical
issues associated with such products.)
  We  
define the {\em entropy rate} of a general source $\Sigma$ as 
\begin{eqnarray}
S(\Sigma) \equiv 
\limsup_{n \rightarrow \infty} \frac{ S(\rho_s^n)}{n}.
\end{eqnarray}

A special case of this general definition of a quantum source
is the i.i.d. source $(H_s,\{\rho_s, \rho_s \otimes \rho_s,...,
\rho_s^{\otimes n},...\}),$ for some fixed $\rho_s.$ Such a source
corresponds to the classical notion of an {\em independent, identically
distributed} classical source, thus the term i.i.d.  The entropy rate 
of this source is simply $S(\rho_s).$

A {\em discrete memoryless channel}, $(H_c,{\cal N})$ consists of a
finite-dimensional Hilbert space, 
$H_c$, and a trace-preserving quantum operation
${\cal N}$. The {\em $n$-th extension} of that channel is given by the
pair $(H_s^{\otimes n},{\cal N}^{\otimes n})$, where $\otimes n$ is used to
denote $n$-fold tensor products.  The memoryless nature of the 
channel is reflected in the fact that the operation performed
on the $n$ copies of the channel system is a tensor product of
independent single-system operations.

Define an {\em $n$-code} $({\cal C},{\cal D})$ from $H_s$ into $H_c$ to consist of a
trace-preserving quantum operation, ${\cal C}$,
from $H_s^{\otimes n}$ to $H_c^{\otimes n}$, and a trace-preserving
quantum operation ${\cal D}$ from $H_c^{\otimes n}$ to
$H_s^{\otimes n}$. We  refer to ${\cal C}$ as the {\em encoding}
and ${\cal D}$ as the {\em decoding}.

The {\em total coding operation} ${\cal T}$ is given by
\begin{eqnarray}
{\cal T} \equiv {\cal D} \circ {\cal N}^{\otimes n} \circ {\cal C}.
\end{eqnarray}
The measure of success we use for the total procedure
is the {\em total entanglement fidelity},
\begin{eqnarray}
F_e(\rho_s^n,{\cal T}). \end{eqnarray}

In practice we frequently abuse notation, usually by omitting
explicit mention of the Hilbert spaces $H_s$ and $H_c$. Note also that
in principle the channel could have different input and output Hilbert
spaces. To ease notational clutter we do not consider that case here,
but all the results we prove go through without change.

Given a source state $\rho_s$ and a channel ${\cal N}$, the goal of
noisy channel coding is to find an encoding ${\cal C}$ and
a decoding ${\cal D}$ such that $F_e(\rho_s,{\cal T})$ is close to
one; that is, $\rho_s$ and its entanglement is transmitted almost
perfectly. In general this is not possible to do. However, Shannon
showed in the classical context that by considering blocks of output
from the source, and performing block encoding and decoding it is possible to considerably expand the class of source states
$\rho_s$ for which this is possible. The quantum mechanical 
version of this procedure is to find a sequence
of $n$-codes, $({\cal C}^n,{\cal D}^n)$ such that as
$n \rightarrow \infty$, the measure of success
$F_e(\rho_s^n,{\cal T}^n)$ approaches one, where
${\cal T}^n = {\cal D}^n \circ {\cal N}^{\otimes n} \circ {\cal C}^n$.
(We will sometimes refer to such a sequence as a 
{\em coding scheme}.)

Suppose such a sequence of codes exists for a given source $\Sigma.$
In this case the channel is said to transmit $\Sigma$ 
reliably.
We also say that the channel can transmit reliably at a {\em rate} 
$R = S(\Sigma).$  (Note that this definition does not require that
the channel be able to transmit reliably {\em any} source with 
entropy rate less than or equal to $R$;  that is a different
potential definition of what it means for a channel to transmit
reliably at rate $R$.  
We conjecture that the two definitions are equivalent in the contexts
considered in this paper.)

A noisy channel coding theorem enables one to determine,
for any source and channel, whether or not the source can be transmitted reliably on that channel.  Classically, this is determined by
comparing the entropy rate of the source to the capacity of the
channel.  If the entropy rate of the source
is greater than the capacity, the source cannot be
transmitted reliably.  If the entropy rate is less
than the capacity, it 
can.  The conjunction of these two statements is precisely the 
noisy channel coding theorem.  (The case when the entropy rate of the
source equals the capacity requires separate
consideration;  sometimes reliable transmission is achievable, and
sometimes not.)  We expect that in quantum
mechanics, the entropy rate $S(\Sigma)$ of the source will play the
role of the classical entropy rate.   
A channel will be able to transmit reliably any source with
entropy rate less than
the capacity;  furthermore, {\it no} source with entropy
rate greater than the
capacity will be reliably transmissible (i.e., the channel will be
unable to transmit reliably at a rate  
greater than the capacity.)  The first part of this would  constitute 
a quantum noisy channel coding theorem;  the second, a ``weak 
converse'' of the theorem.  (A ``strong converse'' would require not
just that no source with entropy rate greater than the capacity can be reliably 
transmitted, i.e. transmitted with asymptotic fidelity approaching
unity, but would require that all such sources have asymptotic
fidelity of transmission approaching zero.)

\section{Upper Bounds on the Channel Capacity}
\label{sect: upper bounds}
In this section we investigate a variety of upper bounds on the
capacity of a noisy quantum channel. 
\subsection{Unitary Encodings}
\label{subsect: unitary encodings}

This subsection is concerned with the case where the encoding, ${\cal C}$,
is unitary.

For this subsection only we define
\begin{eqnarray}
C_n \equiv \max_{\rho} I(\rho,{\cal N}^{\otimes n}), \end{eqnarray}
where the maximization is over all inputs $\rho$ to $n$ copies of the channel.
The bound on the channel capacity proved in this section is defined by
\begin{eqnarray} \label{eqtn: unitary bound}
C({\cal N}) \equiv \lim_{n \rightarrow \infty} \frac{C_n}{n}.
\end{eqnarray}
It is not immediately obvious that this limit exists. To see that it
does, notice that $C_n \leq n \log_2 d$
and $C_m + C_n \leq C_{m+n}$ and
apply the lemma proved in Appendix \ref{appendix: limit lemma}. Notice
that $C({\cal N})$ is a function of the channel operation only.

We begin with a theorem that places a limit on the entropy rate of a source which
can be sent through a quantum channel.

{\em Theorem}

Suppose we consider a source $\Sigma= (H_s, \{..\rho_s^n...\})$ 
and a sequence of unitary encodings ${\cal U}^n$ for the
source. Suppose further that there exists
a sequence of decodings, ${\cal D}^n,$ such that
\begin{eqnarray}
\lim_{n \rightarrow \infty} F_e(\rho_s^n,{\cal D}^n \circ {\cal N}^{\otimes n} 
	\circ {\cal U}^n) = 1.
\end{eqnarray}
Then
\begin{eqnarray}
S(\Sigma) \equiv \limsup_{n \rightarrow \infty} \frac{ S(\rho_s^n)}{n} \leq C({\cal N}).
\end{eqnarray}

This theorem tells us that we cannot reliably transmit more than
$C({\cal N})$ qubits of information per use of the channel. 

For unitary ${\cal U}^n$ we have
\begin{eqnarray}
I(\rho_s,{\cal N}^{\otimes n} \circ {\cal U}^n) =
	 I({\cal U}^n(\rho_s),{\cal N}^{\otimes n}),
\end{eqnarray}
and thus
\begin{eqnarray}
I(\rho_s,{\cal N}^{\otimes n} \circ {\cal U}^n) \leq C_n. \end{eqnarray}
By (\ref{eqtn: fidelity lemma}) with ${\cal E} \equiv {\cal N}^{\otimes n}
\circ {\cal U}^n$, and the fact
that $I({\cal U}^n(\rho_s),{\cal N}^{\otimes n}) \leq
\max_{\rho} I(\rho,{\cal N}^{\otimes n}) \equiv C_n$, it now follows that
\begin{eqnarray}
\frac{S(\rho_s^n)}{n} & \leq & \frac{C_n}{n} + \frac{2}{n} + \nonumber \\
 & & 4 (1-F_e(\rho_s^n,{\cal D}^n \circ {\cal N}^{\otimes n} \circ {\cal U}^n))
	\log d. \end{eqnarray} (Note that $d$ here is the dimension 
of a single copy of the source Hilbert space, so that we have inserted
$d^n$ for the overall dimension $d$ of (\ref{eqtn: fidelity lemma})). 
Taking $\limsup$s on both sides of the equation completes the proof of the
theorem.





It is extremely useful to study this result at length, since the basic
techniques employed to prove the bound are the same as those that
appear in a more elaborate guise later in the paper.  
In particular, what features of quantum mechanics
necessitate a change in the proof methods used to obtain the classical
bound?

Suppose the quantum analogue of the classical subadditivity of mutual
information were true, namely
\begin{eqnarray}
I(\rho^n,{\cal N}^{\otimes n}) \leq \sum_{i=1}^n I(\rho^n_i,{\cal N}),
\end{eqnarray}
where $\rho^n$ is any density operator that can be used as input to $n$
copies of the channel, and $\rho^n_i$ is the density operator obtained
by tracing out all but the $i$-th channel. Then 
it would follow easily from the definition that $C_n = C_1$ for all
$n$, and thus 
\begin{eqnarray}
C({\cal N}) = C_1 = \max_{\rho} I(\rho,{\cal N}). \end{eqnarray} 
This expression is exactly analogous to the classical expression for
channel capacity as a maximum over input distributions of the mutual
information between channel input and output. If this were truly a
bound on the quantum channel capacity then it would allow easy
numerical evaluations of bounds on the channel capacity, as the
maximization involved is easy to do numerically, and the coherent
information is not difficult to evaluate.  

Unfortunately, it is not possible to assume that the quantum
mechanical coherent information is subadditive, as shown by example
(\ref{eqtn: example 2}), and thus in general it is possible that
\begin{eqnarray}
C({\cal N}) > C_1. \end{eqnarray}
We will later discuss results of Shor and Smolin \cite{Shor96a} which
demonstrate that the channel capacity can exceed $C_1$. 

Notice that to evaluate the bound $C({\cal N})$ involves
taking the limit in (\ref{eqtn: unitary bound}). 
To numerically evaluate this limit directly is certainly not a
trivial task, in general. The result we have presented, that
(\ref{eqtn: unitary bound}) is an upper bound on channel capacity, is
an important theoretical result, that may aid in the development of
effective numerical procedures for obtaining general bounds. But it
does not yet constitute an effective procedure.

\subsection{General Encodings}
\label{sect: general encodings}

We now consider the case where something more general than a
unitary encoding is allowed. In principle, it is always possible to
perform a non-unitary encoding, ${\cal C}$, by introducing
an extra ancilla system, performing a joint unitary on the source
plus ancilla, and then discarding the ancilla.

We define
\begin{eqnarray}
C_n \equiv \max_{\rho,{\cal C}} I(\rho,{\cal N}^{\otimes n}\circ {\cal C}),
\end{eqnarray}
where the maximization is over all inputs $\rho$ to the encoding operation,
${\cal C}$, which in turn maps to $n$ copies of the channel.
The bound on the channel capacity proved in this section is defined by
\begin{eqnarray} \label{eqtn: general bound}
C({\cal N}) \equiv \lim_{n \rightarrow \infty} \frac{C_n}{n}.
\end{eqnarray}
Once again, to prove that this limit exists one applies the lemma
proved in Appendix \ref{appendix: limit lemma}.

To prove that this quantity is a bound on the channel capacity, one
applies almost exactly the same reasoning as in the preceding subsection.
The result is:

{\em Theorem}
Suppose we consider a source $\Sigma= (H_s, \{...\rho_s^n...\})$ 
and a sequence of encodings ${\cal C}^n$ for the
source. Suppose further that there exists
a sequence of decodings, ${\cal D}^n$ such that
\begin{eqnarray}
\lim_{n \rightarrow \infty} F_e(\rho_s^n,{\cal D}^n \circ {\cal N}^{\otimes n} 
	\circ {\cal C}^n) = 1.
\end{eqnarray}
Then
\begin{eqnarray}
S(\Sigma) \equiv \limsup_{n \rightarrow \infty} \frac{ S(\rho_s^n)}{n} \leq C({\cal N}).
\end{eqnarray}

Again, this result places an upper bound on the rate at which information
can be reliably transmitted through a noisy quantum channel. 
The proof is very
similar to the earlier proof of a bound for unitary encodings. One simply
applies (\ref{eqtn: fidelity lemma}) with
${\cal E} = {\cal N}^{\otimes n} \circ {\cal C}^n$ and ${\cal D} = {\cal D}^n,$  to give:
\begin{eqnarray}
\frac{S(\rho_s^n)}{n} & \leq & \frac{C_n}{n} + \frac{2}{n} +
	4(1-F_e(\rho_s^n,{\cal D}^n \circ {\cal N}^{\otimes n} \circ 
	{\cal C}^n)) \log_2 d. \nonumber \\
& & 
\end{eqnarray}
Taking $\limsup$s on both sides of the equation completes the proof.

It is instructive to see why the proof fails when the maximization is done
over channel input states alone, rather than over all source states and
encoding schemes. The basic idea is that there may exist source states,
$\rho_s$, and encoding schemes ${\cal C}$, for which
\begin{eqnarray} \label{eqtn: strange inequality}
I(\rho,{\cal N} \circ {\cal C}) > I({\cal C}(\rho),{\cal N}).
\end{eqnarray}
This possibility stems from the failure of the quantum pipelining
inequality, (\ref{eqtn: example 1}). 
It is clear that the existence of such a scheme would cause the line of
proof suggested above to fail.  Classically the pipelining inequality
holds, and therefore the complication of having to
maximize over encodings does not arise.

Having proved that $C({\cal N})$ is an upper bound on the channel capacity,
let us now investigate some of the properties of this bound. First
we examine the range over which $C({\cal N})$ can vary. Note that
\begin{eqnarray}
0 \leq C_n \leq n \log_2 d, \end{eqnarray}
since if $\rho$ is pure then 
$I(\rho,{\cal N}^{\otimes n} \circ {\cal C}) = 0$
for any encoding ${\cal C}$, and for all $\rho$ and ${\cal C}$,
$I(\rho,{\cal N}^{\otimes n} \circ {\cal C}) \leq \log_2 d^n = n
	 \log_2 d$, 
since the
channel output has $d^n$ dimensions. It follows that
\begin{eqnarray}
0 \leq C({\cal N}) \leq \log_2 d. \end{eqnarray}
This parallels the classical result, which states that
the channel capacity varies between $0$ and $\log_2 s$,
where $s$ is the
number of channel symbols. The upper bound on the classical capacity
is attained if and only if the classical channel is noiseless. 

In the case when ${\cal N}$ takes a constant value,
\begin{eqnarray}
{\cal N}(\rho) = \sigma, \end{eqnarray}
for all channel inputs, $\rho$, it is not difficult to verify that 
$C({\cal N}) = 0$. This is consistent with the obvious fact that the
capacity for quantum information of such a channel is zero.

The ``completely decohering channel'' is defined by 
\begin{eqnarray}
{\cal N}(\rho) =
\sum_i P_i \rho P_i,
\end{eqnarray}
with $P_i \equiv |i \rangle \langle i|$ 
a complete orthonormal set of one-dimensional projectors. 
This channel is classically noiseless, yet a straightforward 
application of (\ref{eqtn: abstract subadditivity}) yields
$C({\cal N}) = 0,$ and therefore this channel has zero capacity
for the transmission of entanglement.  

More generally, if ${\cal N}(\rho) = \sum_i A_i \rho A_i^{\dagger}$, where
$A_i = \lambda_i |a_i\rangle \langle b_i|$, then $C({\cal N}) = 0$
by the same argument, and thus the channel capacity for such a channel
is zero. As a special case of this result, it follows that the capacity
of {\em any} classical channel as defined in section
\ref{sect: classical} to transmit entanglement is zero.

Provided the input and output dimensions of the channel are the same,
it is not difficult to show that  $C({\cal N}) = \log_2 d$ if
and only if ${\cal N}$ is unitary. 

It is also of interest to consider what happens when channels ${\cal N}_1$
and ${\cal N}_2$ are composed, forming a concatenated channel,
${\cal N} = {\cal N}_2 \circ {\cal N}_1$. From the data processing
inequality it follows that
\begin{eqnarray} \label{eqtn: Albert Einstein}
C({\cal N}_1) \geq C({\cal N}). \end{eqnarray}
It is clear by repeated application of the data-processing inequality
that this result also holds if we compose more than two
channels together, and even holds if we allow intermediate decoding and
re-encoding stages. Classical channel capacities also behave in
this way: the capacity of
a channel made by composing two (or more) channels together is no greater
than the capacity of the first part of the channel alone.

Although (\ref{eqtn: example 1}) might seem to suggest otherwise, 
in fact
\begin{eqnarray} \label{eqtn: Ahristopher Fuchs}
C({\cal N}_2) \geq C({\cal N}). \end{eqnarray}
For let us suppose that ${\cal C}$ is the encoding which achieves
$C({\cal N}),$ so that the total operation 
is ${\cal D} \circ {\cal N} \circ {\cal C} \equiv {\cal D} \circ {\cal N}_2 \circ {\cal N}_1 \circ {\cal C}.$   As our encoding for the channel 
${\cal N}_2$, we may use ${\cal N}_1 \circ {\cal C}$ and decode with
${\cal D},$ hence achieving
precisely the same total operation.

Inequalities analogous to (\ref{eqtn: Albert Einstein}) 
and (\ref{eqtn: Ahristopher Fuchs}) may also be stated 
for the actual channel capacity.  Clearly
these statements are true as well.

\subsection{Other Encoding Protocols}
\label{subsect: other encoding protocols}

So far we have considered two allowed classes of encodings: encodings where
a general unitary operation can be performed on a block of quantum systems, and
encodings where a general trace-preserving quantum operation can be
performed on a block of quantum systems. If large-scale quantum computation
ever becomes feasible it may be realistic to consider encoding protocols
of this sort. However, for present-day applications of quantum
communication such as quantum cryptography and teleportation, 
it is realistic to consider much more restricted classes of encodings.
In this section we describe several such classes.

We begin by considering the class involving local unitary operations only. 
We refer to this class as $U$-$L$. It consists of the set of
operations ${\cal C}$ which can be written in the form
\begin{eqnarray} \label{eqtn: local unitary encodings}
{\cal C}(\rho) = (U_1 \otimes \cdots \otimes U_n) \rho (U_1^{\dagger} \otimes
	\cdots \otimes U_n^{\dagger}), \end{eqnarray}
where $U_1,\ldots,U_n$ are local unitary operations on systems $1$ through
$n$.
Another possibility is the class $L$ of encodings involving
local operations only, i.e. operations of the form:
\begin{eqnarray} \label{eqtn: local encodings}
\sum_{i_1,...,i_n} (A_{i_1} \otimes B_{i_2} \otimes \cdots
\otimes Z_{i_n}) \rho \nonumber & & \\
 (A_{i_1}^\dagger \otimes B_{i_2}^\dagger \otimes \cdots
\otimes Z_{i_n}^\dagger). 
\end{eqnarray}
In other words, the overall operation has a tensor product form
${\cal A} \otimes {\cal B} \otimes \cdots \otimes {\cal Z}$.

A more realistic class is $1$-$L$ -- encoding by local
operations with one way classical communication. The idea is that the encoder
is allowed to do encoding by performing arbitrary quantum operations
on individual members (typically, a single qubit) of the strings
of quantum systems emitted by a source. This is not unrealistic with present
day technology for manipulating single qubits. Such operations could
include arbitrary unitary rotations, and also generalized measurements. After
the qubit is encoded, the results of any measurements done during the
encoding may be used to assist in the encoding of later qubits. This
is what we mean by one way communication - the results of the measurement
can only be used to assist in the encoding of later qubits, not earlier
qubits.

Another possible class is $2$-$L$ - encoding by local
operations with two-way classical communication. This may arise
in a situation where there are many identical channels operating
side by side in space. Once again it is assumed
that the encoder can perform arbitrary local operations, only this time
two-way classical communication is allowed when performing the encoding.

For any class of encodings $\Lambda$ arguments analogous to 
those used above for general and for unitary block coding, ensure
that the expression 
\begin{eqnarray}
C_{\Lambda}({\cal N}) \equiv \lim_{n \rightarrow \infty} \frac{C_{\Lambda,n}}{n},
\end{eqnarray}
where
\begin{eqnarray}
C_{\Lambda,n} \equiv \max_{\rho,{\cal C} \in \Lambda} 
	I(\rho,{\cal N}^{\otimes n} \circ {\cal C}),
\end{eqnarray}
is an upper bound to the rate at which quantum information can be
reliably transmitted using encodings in $\Lambda$.  Thus, in addition to the
bounds for general and unitary encodings, there are bounds
$C_{U-L}, C_{L}, C_{1-L},$ and $C_{2-L},$
which provide upper bounds on the rate of quantum information
transmission for these types of encodings. {A priori} it is not clear what
the exact relationships are amongst these bounds, although various inequalities
may easily be proved,
\begin{eqnarray}
C_{U-L} \leq C_{L} \leq C_{1-L} \leq C_{2-L} \leq C_{\mbox{general}} \\
C_{U-L} \leq C_{\mbox{unitary}} \\
C_{\mbox{unitary}} \leq C_{\mbox{general}}.
\end{eqnarray}
Furthermore, note that these bounds allow general decoding schemes. It is
possible that much tighter bounds may result if we restrict the decoding schemes
in the same way we have restricted the encoding schemes.

An interesting and important question is whether there are
closed-form
characterizations of the sets of quantum operations corresponding to
particular types of encoding schemes such as $1$-$L$ and $2$-$L$. For
example, in the cases of $U$-$L$ and $L$ there are explicit forms 
(\ref{eqtn: local unitary encodings},\ref{eqtn: local encodings}) for the classes of encodings
allowed.  For $1$-$L$ the operations take the form:
\begin{eqnarray} \label{eqtn: 1-L explicit}
\sum_{i_1,...i_n} (A_{i_1} &\otimes& B_{i_1,i_2} \otimes \cdots
\otimes Z_{i_1,i_2,...i_n}) \rho \nonumber \\
&&(A_{i_1}^\dagger \otimes B_{i_1,i_2}^\dagger \otimes \cdots
\otimes Z_{i_1,i_2,...i_n}^\dagger).
\end{eqnarray}
A drawback to this expression is that it is not written in a closed form,
making it difficult to perform optimizations over $1$-$L$.
It would be extremely valuable to obtain a closed form for the set of operations in
$1$-$L$. One possible approach to doing this is
to limit the range of the indices in the previous
expression.  This is related to the number of rounds
of classical communication which are involved in the operation.
Similar remarks to these also apply to the class $2$-$L$.
Indeed, it is not yet clear to us if there is an expression analogous
to (\ref{eqtn: 1-L explicit})for $2$-$L$ encodings.  One possibility is:
\begin{eqnarray}
\sum_{i} (A_{i} \otimes B_{i} \otimes \cdots
\otimes Z_{i}) \rho (A_{i}^\dagger \otimes B_{i}^\dagger \otimes \cdots
\otimes Z_{i}^\dagger).
\end{eqnarray}
 However, although all $2$-$L$ operations involving
a finite number of rounds of communication can 
certainly be put in this form, we do not presently see why all operations
expressible in this form should be realizable with local operations
and two-way classical communication.

The classes we have described in this subsection are certainly
not the only realistic classes of encodings. Many
more classes may be considered, and in specific applications this
may well be of great interest. What we have done is
illustrated a general technique for
obtaining {\em bounds} on the channel capacity for different classes of
encodings. A major difference between classical information theory
and quantum information theory is the greater interest in the quantum
case in studying different classes of encodings. Classically it is, in
principle, possible to perform an arbitrary encoding and decoding operation
using a look-up table. However, quantum
mechanically this is far from being the case, so there
is correspondingly more interest in studying
the channel capacities that may result from considering different classes
of encodings and decodings. 

\section{Discussion}
\label{sect: discussion}

What then can be said about the status of the quantum
noisy channel coding theorem in the light of comments made in
the preceding sections?  While we have established upper bounds,
we have not proved achievability of these bounds. How might
one prove that these bounds are achievable?

Lloyd \cite{Lloyd97a} has also
proposed an expression involving a maximum of the
coherent information as the channel 
capacity, 
\begin{eqnarray} \label{eqtn: Lloyd}
\max_{\rho} I(\rho,{\cal N}),
\end{eqnarray}
and outlines a technique involving random coding for
achieving rates up to this quantity.
The criterion for reliable transmission
used by Lloyd appears to be the subspace fidelity
criterion of eqtn. (\ref{eqtn: subspace fidelity}). As noted
earlier, this criterion is at least as strong as the
criterion based on entanglement fidelity which we have been
using, that is, 
asymptotically good coding schemes with respect to subspace
fidelity are also asymptotically good with respect to the
entanglement fidelity.

Suppose one applies coding schemes to achieve rates up to
(\ref{eqtn: Lloyd}), but with the basic system used in blocking
taken to be $n$ of the old systems. Then
it is clear that rates up to
\begin{eqnarray} \label{eqtn: unitary bound 2}
\max_{\rho} \frac{I(\rho,{\cal N}^{\otimes n})}{n}
\end{eqnarray}
may be achieved using such coding schemes, where the maximization is
done over density operators for $n$ copies of the source. It
follows that rates up to
\begin{eqnarray}
\lim_{n \rightarrow \infty} \max_{\rho}
	\frac{I(\rho,{\cal N}^{\otimes n})}{n} \end{eqnarray}
may be achieved. This quantity is simply the bound
(\ref{eqtn: unitary bound}) which we found earlier for
noisy channels with the class of encodings restricted to be
unitary. As remarked in the last section, it is in
general not possible to 
identify the quantity appearing in the previous equation
with the quantity (\ref{eqtn: Lloyd}), because the 
coherent information is not, in general,
subadditive, cf. eqtn. (\ref{eqtn: example 2}).

The coding schemes considered by Lloyd appear to be
restricted to be
projections followed by unitaries. We call such 
encodings {\em restricted encodings}, since they do not cover
the full class of encodings possible. For the purposes of
proving upper bounds it is not sufficient to consider
a restricted
class of encodings, since it is possible that
other coding schemes may do better, and therefore
that the capacity is somewhat larger than
(\ref{eqtn: unitary bound 2}).
We suspect that this is not the case, but have been
unable to provide a rigorous proof. A heuristic argument
is provided in subsection \ref{subsect: heuristics}.

In the light of these remarks it is interesting that
the coding scheme of Shor and Smolin
\cite{Shor96a} provides an example where rates of
transmission exceeding (\ref{eqtn: Lloyd})
are obtained. Nevertheless, the general bound
(\ref{eqtn: general bound}) must still be obeyed by
a coding scheme of their type.

However, one can still make progress towards a proof that the
expression, (\ref{eqtn: general bound}), which
bounds the channel capacity, is the correct capacity.
If we accept that it is possible to attain rates up to
(\ref{eqtn: unitary bound 2}),
then the following four-stage construction shows
that (\ref{eqtn: general bound}) is a
correct expression for the capacity; i.e. that in addition
to being an upper bound as shown in
section \ref{sect: upper bounds}, it is 
also achievable.
\begin{figure}
\epsfxsize 3.4in
\epsfbox{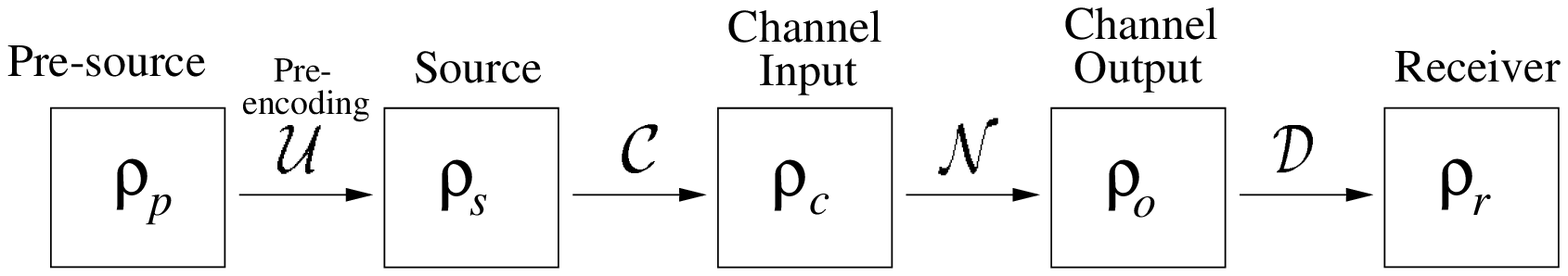}
\caption{} \label{fig: channel3}
Noisy quantum channel with an extra stage, a restricted {\em pre-encoding},
${\cal U}$.
\end{figure}

For a fixed block size, $n$, one finds an
encoding, ${\cal C}^n$, for which the maximum in
\begin{eqnarray}
C_n \equiv \max_{{\cal C}^n,\rho_s} I(\rho_s,{\cal C}^n),
\end{eqnarray}
is achieved. One then regards the composition
${\cal N}^{\otimes n} \circ {\cal C}^n$ as a single
noisy quantum channel,
and applies the achievability
result on restricted encodings to the joint channel
${\cal N}^{\otimes n} \circ {\cal C}^n$ to achieve an even
longer $mn$ block coding scheme with high entanglement
fidelity.

This gives a joint coding scheme
${\cal U}^{mn} \circ ({{\cal C}^n})^{\otimes m}$ which
for sufficiently large blocks $m$ and $n$ can come
arbitrarily close to
achieving the channel capacity (\ref{eqtn: general bound}).
An important open question is
whether (\ref{eqtn: general bound})
is equal to (\ref{eqtn: unitary bound}). It is clear that
the former expression is at 
least as large as the latter; we give a heuristic argument
for equality in the next subsection, but rigorous results are needed.

Thus, we think it likely that the
expression (\ref{eqtn: unitary bound}) will turn out to 
be the maximum achievable rate of reliable
transmission through 
a quantum channel.  But this is still not satisfactory as
an expression for the capacity, because of the difficulty
of evaluating the limit involved.  At a minimum,
we would like to 
know enough about the rate of convergence
of $C_n$ to its limit
to be able to accurately estimate the
error in a numerical calculation
of capacity, thus providing an effective procedure for
calculating the capacity to any desired degree of
accuracy.  It would also be useful to know
whether the coherent information 
is concave in the input density operator;
if this were so, it
would greatly aid in establishing that one has reached a 
maximum, since in this case a local maximum is guaranteed to 
be a global maximum (Appendix \ref{appendix: convexity}).  

\subsection{Unitary versus Nonunitary Encoding}
\label{subsect: heuristics} 

For the purposes of obtaining a capacity
theorem for general encodings and decodings, a restriction on
the class of encodings is clearly unacceptable.  For
example, given a source density operator whose
eigenvalues are not all
equal, we may not even be able to send it
reliably through a noiseless channel
whose capacity is just greater than the
source entropy rate without doing non-unitary
compression as described
in \cite{Schumacher95,Jozsa94a,Barnum96}.  This
compression, which is essentially projection onto the
{\em typical subspace} \cite{Schumacher95} of
the source, is not a unitary operation, and thus we expect 
that nonunitary operations will be essential to showing achievability
of the noisy channel capacity.


We conjecture that once the projection onto the typical source
subspace is 
accomplished, nonunitary operations are of no further use in 
achieving reliable transmission through a noisy channel.  Although
we have not yet rigorously shown this, we give a heuristic argument
below. If the conjecture is true then it can
be used to show that expressions
(\ref{eqtn: unitary bound}) and (\ref{eqtn: general bound})
are equal.


Our heuristic argument applies only to sources for which
a {\em typical subspace} \cite{Schumacher95} exists.
This includes
all i.i.d. sources, for which the output is of the form
$\rho_s^{\otimes n}$. Let $\Lambda$ be the projector onto
the typical subspace after $n$ uses of the source, and
$\overline \Lambda$ the projector onto the
orthogonal subspace.
Given any positive $\delta$ it is true that for
sufficiently large $n$,
\begin{eqnarray}
\mbox{tr}(\overline \Lambda \rho_s^{\otimes n} \overline \Lambda )
	\leq \delta.
\end{eqnarray}
Defining the restriction of the source to the typical subspace,
\begin{eqnarray}
\rho^n_T \equiv \frac{\Lambda \rho_s^{\otimes n} \Lambda}{
	\mbox{tr}(\Lambda \rho_s^{\otimes n} \Lambda)},
\end{eqnarray}
and applying the continuity lemma for entanglement fidelity,
(\ref{eqtn: continuity lemma}), we see that
\begin{eqnarray}
|F_e(\rho_T^n,{\cal E})-F_e(\rho_s^{\otimes n},{\cal E}) |
\leq
\frac{4 \delta}{(1-\delta)^2}, \end{eqnarray}
for any trace-preserving operation ${\cal E}$.
By choosing $n$ sufficiently large $\delta$ can be
made arbitrarily small, and thus we see that
for the entanglement fidelity for the source to be
high asymptotically, it is necessary and sufficient that the
entanglement fidelity be high asymptotically for
the restriction of the source to the typical
subspace.

We now come to the heuristic argument.
In order that the entanglement fidelity for the total channel
be high, it is asymptotically
necessary and sufficient that 
the composite operation
${\cal D}^n \circ {\cal N}^{\otimes n} \circ {\cal C}^n$
have high entanglement fidelity when the source is
restricted to the typical subspace, $\tau$.
Hence, if an encoding ${\cal C}^n$ is nonunitary 
on $\tau,$ it must be ``close to reversible'' on
$\tau,$ and ${\cal D}^n \circ {\cal N}^{\otimes n}$ must
be close to reversing it.
In \cite{Nielsen97a} it is shown that perfect
reversibility of an 
operation on a subspace $M$ is equivalent to
the statement that the
operation, restricted to that subspace, may be represented by
operators
$\{\sqrt{p_i} U_i P_M \},$ where
$P_M U _j^\dagger U_i P_M = \delta_{ij} P_M$
and $P_M$ is the projector onto $M$. 
That is, the operation randomly
(with probabilities $p_i$) chooses a 
unitary which moves the state into one
of a mutually orthogonal set
of subspaces.  Hence ${\cal C}^n,$ {\em in its action on the
source's typical subspace,} is close
to some perfectly reversible operation ${\cal C}_*^n$
consisting of ``randomly picking a unitary into an 
orthogonal subspace.''  Hence the entanglement fidelity of
the total operation
${\cal T}$ is close to that of ${\cal T}_*^n,$ in which
the encoding
${\cal C}^n$ is replaced with ${\cal C}^n_*.$
The linearity of the entanglement fidelity in the operation
implies that for at least one of the unitaries $U_i$ in the 
random-unitaries representation of the perfectly reversible
operation ${\cal C}_*^n$, the entanglement fidelity is at least as
good if the unitary is substituted for ${\cal C}^n_*$.
Therefore, arbitrary encodings ${\cal C}^n$ are
close to unitary encodings of $\tau$ into a subspace of
the channel's Hilbert space.  Thus the only nonunitarity
which it is necessary to consider is the restriction
to the source's typical subspace. 

\section{Channels with a Classical Observer}
\label{sect: observed channel}

In this section we consider a generalized version of the quantum
noisy channel coding problem. Suppose that in addition to a noisy
interaction with the environment there is also
a classical observer who is able to perform a measurement. This measurement
may be on the channel or the environment of the channel, or possibly on
both.

\begin{figure}
\epsfxsize 3.4in
\epsfbox{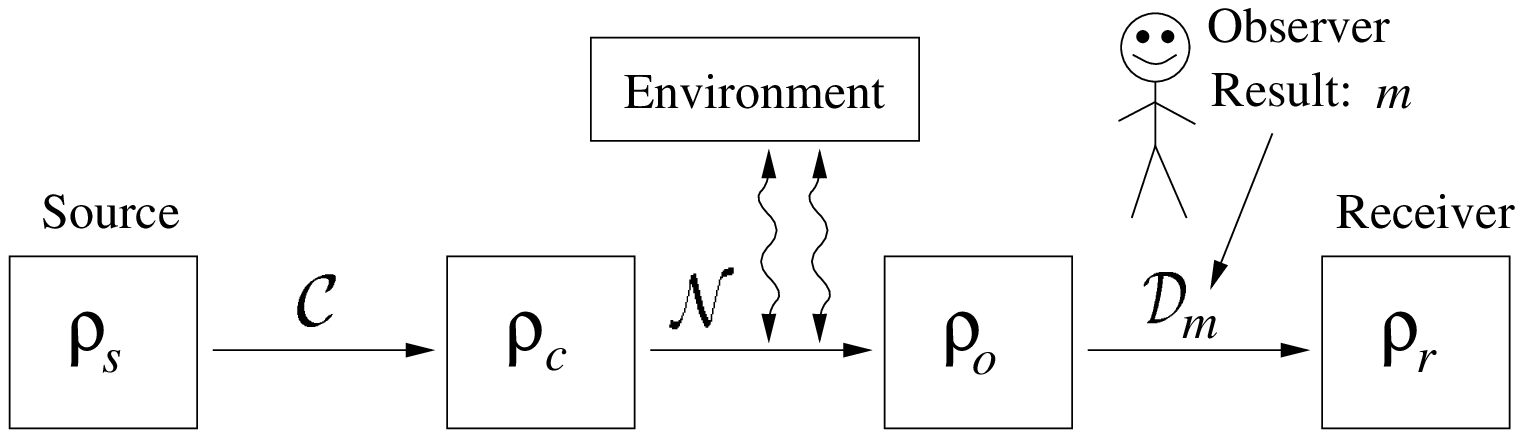}
\caption {} \label{fig: channel4}
Noisy quantum channel with a classical observer.
\end{figure}

The result of the measurement is then sent to the decoder, who may use the
result to assist in decoding. We assume that this
transmission of classical information is done noiselessly, although it
is also interesting to consider what happens when the classical
transmission also involves
noise. It can be shown \cite{Kraus83a} that the state received by the
decoder is again related to the state $\rho$ used as input to the
channel by a quantum operation ${\cal N}_m$, where $m$ is the
measurement result recorded by the classical observer,
\begin{eqnarray}
\rho \rightarrow \frac{{\cal N}_m(\rho)}{\mbox{tr}({\cal N}_m(\rho))}.
\end{eqnarray}
The basic situation is illustrated in figure \ref{fig: channel4}. The idea
is that by giving the decoder access to classical information about
the environment responsible for noise in the channel it may be possible
to improve the capacity of that channel, by allowing the decoder
to choose different decodings ${\cal D}_m$ depending on the measurement
result $m$.

We now give a simple example which illustrates that this can be the case. 
Suppose we have a two-level system in a state $\rho$ and
an initially uncorrelated four-level environment initially in
the maximally mixed state $I/4$, so the total state of the joint system is
\begin{eqnarray}
\rho \otimes \frac{I}{4}.
\end{eqnarray}
We fix an orthonormal basis $|1\rangle,|2\rangle,|3\rangle,
|4\rangle$ for the environment. We assume that a unitary interaction
between the system and environment takes place, given by the
unitary operator
\begin{eqnarray}
U & = & I \otimes |1\rangle \langle 1| + \sigma_x \otimes |2\rangle
	\langle 2| + \nonumber \\
	& & \sigma_y \otimes |3\rangle \langle 3| +
	\sigma_z \otimes |4 \rangle \langle 4|.
\end{eqnarray}
The output of the channel is thus
\begin{eqnarray}
\rho \rightarrow {\cal N}(\rho) \equiv \mbox{tr}_E \left( U \left( \rho \otimes
	\frac{I}{4} \right) U^{\dagger} \right).
\end{eqnarray}
The quantum operation ${\cal N}$ can be given two particularly
useful forms,
\begin{eqnarray}
{\cal N}(\rho) & = & \frac{1}{4} \left( I\rho I + \sigma_x \rho \sigma_x
	+ \sigma_y \rho \sigma_y + \sigma_z \rho \sigma_z \right) \\
	& = & \frac{I}{2}.
\end{eqnarray}
It is not difficult to show from the second form that
\begin{eqnarray}
C({\cal N}) = 0,
\end{eqnarray}
and thus the channel capacity for the channel ${\cal N}$ is equal to zero.
Suppose now that an observer is introduced, who is allowed to perform
a measurement on the environment. This measurement is
a Von Neumann measurement in the $|1\rangle,|2\rangle,|3\rangle,|4\rangle$
basis, and yields a corresponding measurement result, $m = 1,2,3,4$.
Then the quantum operations corresponding to these four measurement outcomes
are
\begin{eqnarray}
{\cal N}_1(\rho) & = & \frac 14 \rho \\
{\cal N}_2(\rho) & = & \frac 14 \sigma_x \rho \sigma_x \\
{\cal N}_3(\rho) & = & \frac 14 \sigma_y \rho \sigma_y \\
{\cal N}_4(\rho) & = & \frac 14 \sigma_z \rho \sigma_z.
\end{eqnarray}
Each of these is unitary, up to a constant multiplying factor, so conditioned
on knowing the measurement result $m$, the corresponding channel capacity
$C_m$ is perfect. That is,
\begin{eqnarray}
C_m = 1
\end{eqnarray}
for all measurement outcomes $m$.
This is an example where the capacity of the observed channel is strictly greater
than for the unobserved channel.

This result is particularly clear in the context of teleportation. Nielsen
and Caves \cite{Nielsen96b} showed that the problem of teleportation
can be understood as the problem of a quantum noisy channel
with an auxiliary classical channel. In
the single qubit teleportation scheme of
Bennett {\em et al} \cite{Bennett93a} there are four quantum operations
relating the state Alice wishes to teleport to the state Bob
receives, corresponding to each of the four measurement results. In that
scheme it happens that those four operations are the ${\cal N}_m$ we have
described above. Furthermore in the absence of the classical channel, that
is, when Alice does not send the result of her measurement to Bob,
the channel is described by the single operation ${\cal N}$. Clearly, in
order that causality be preserved we expect that the channel capacity
be zero. On the other hand, in order that teleportation be able to occur we
expect that the channel capacity
$C_m = 1$, as was shown above. Teleportation understood
in this way as a noisy channel with a classical side channel offers a
particularly elegant way of seeing that the transmission of quantum
information may sometimes be greatly improved by making use
of classical information.

The remainder of this section is organized into three subsections. Subsection
\ref{subsect: upper bounds on observed channel} proves
bounds on the capacity of an observed channel. This requires nontrivial
extensions of the techniques developed earlier for proving bounds on
the capacity of an unobserved channel. Subsection
\ref{subsect: relationship to unobserved channel} relates work done on
the observed channel to the work done on the
unobserved channel. Subsection \ref{subsect: observed discussion} discusses 
possible extensions to this work on observed channels.

\subsection{Upper Bounds on Channel Capacity}
\label{subsect: upper bounds on observed channel}

We now prove several results bounding the
channel capacity of an observed channel, just as we did earlier for the 
unobserved channel.
The following lemma generalizes the earlier entanglement fidelity lemma
for quantum operations, which was the foundation of our earlier
proofs of upper bounds on the channel capacity.

{\em Lemma} (generalized entanglement fidelity lemma for operations)

Suppose ${\cal E}_m$ are a set of quantum operations such that
$\sum_m {\cal E}_m$ is a trace-preserving quantum operation. Suppose further
that ${\cal D}_m$ is a trace-preserving quantum operation for each $m$.
Then

\begin{eqnarray}
\label{eqtn: fidelity lemma for general operations}
S(\rho) & \leq & \sum_m \mbox{tr}({\cal E}_m(\rho)) I(\rho,{\cal E}_m) + 2 +
	\nonumber \\
 & & 4 (1-F_e(\rho,{\cal T})) \log_2 d,
\end{eqnarray}
where
\begin{eqnarray}
{\cal T} \equiv \sum_m {\cal D}_m \circ {\cal E}_m.
\end{eqnarray}

By the second step of the data processing inequality,
(\ref{eqtn: data processing}),
$I(\rho,{\cal E}_m) \geq I(\rho,{\cal D}_m \circ {\cal E}_m)$ for each $m$,
and noting also that by the trace-preserving property of ${\cal D}_m$,
$\mbox{tr}({\cal E}_m(\rho)) = \mbox{tr}(({\cal D}_m \circ {\cal E}_m)(\rho))$,
we obtain
\begin{eqnarray}
S(\rho) & \leq & S(\rho) + \sum_m \left[ \mbox{tr}({\cal E}_m(\rho))
	I(\rho,{\cal E}_m) - \right. \nonumber \\
	& & \left. \mbox{tr}(({\cal D}_m \circ {\cal E}_m)(\rho))
	I(\rho,{\cal D}_m \circ {\cal E}_m) \right]. \end{eqnarray}
Applying the generalized convexity theorem for coherent information
(\ref{eqtn: abstract subadditivity}) gives
\begin{eqnarray}
-\sum_m \mbox{tr}(({\cal D}_m \circ {\cal E}_m)(\rho)) I(\rho,{\cal D}_m
	\circ {\cal E}_m) & \leq & -I(\rho,{\cal T}). \nonumber \\
& &
\end{eqnarray}
We obtain
\begin{eqnarray}
S(\rho) & \leq & \sum_m \mbox{tr}({\cal E}_m(\rho)) I(\rho,{\cal E}_m)
	+ S(\rho) - I(\rho,{\cal T}).
\end{eqnarray}
But ${\cal T} = \sum_m {\cal D}_m \circ {\cal E}_m$ is trace-preserving since
${\cal D}_m$ is trace-preserving and $\sum_m {\cal E}_m$ is
trace-preserving, and thus by (\ref{eqtn: Araki-Lieb}),
\begin{eqnarray}
S(\rho) - I(\rho,{\cal T}) & = &
	S(\rho) - S({\cal T}(\rho)) + S_e(\rho,{\cal T}) \\
	& \leq & 2 S_e(\rho,{\cal T}).
\end{eqnarray}
Finally, an application of the quantum Fano inequality
(\ref{eqtn: quantum Fano}) along with the observations that the entropy
function $h$ appearing in that inequality is bounded above by one,
and $\log (d^2-1) \leq 2 \log d$, gives
\begin{eqnarray}
S(\rho) & \leq & \sum_m \mbox{tr}(({\cal D}_m \circ {\cal E}_m)(\rho)) 
	I(\rho,{\cal D}_m \circ {\cal E}_m) + 2 + \nonumber \\
 & & 	4(1-F_e(\rho,{\cal T})) \log d,
\end{eqnarray}
as we set out to prove.

If we define 
\begin{eqnarray}
\label{eqtn: observed capacity}
C(\{ {\cal N}_m \}) \equiv  \lim_{n \rightarrow \infty}
	\max_{{\cal C}^n, \rho} \nonumber \\
  \sum_{m_1,...,m_n}  \mbox{tr}((
	{\cal N}_{m_1} \otimes \cdots \otimes {\cal N}_{m_n}
	\circ {\cal C}^n )(\rho)) \nonumber \\  
	\frac{I(\rho, {\cal N}_{m_1} \otimes
	\cdots \otimes {\cal N}_{m_n}\circ {\cal C}^n)}{n},
\end{eqnarray}
we may use (\ref{eqtn: fidelity lemma for general operations}) to easily 
prove that
$C(\{ {\cal N}_m \})$ is an upper bound on the rate of reliable
transmission through an observed channel, in precisely the same way we
earlier used (\ref{eqtn: fidelity lemma}) to prove bounds for
unobserved channels.

We may derive the same bound in another fashion if we associate
observed channels with tracepreserving unobserved channels in the
following fashion suggested by examples in \cite{Bennett97a}.
To an observed channel $\{ {\cal N}_m\}$ we associate a single
tracepreserving operation ${\cal M}$ from $H_c$ to a larger Hilbert
space $H_c \otimes M$, where $M$ is a ``register'' Hilbert space. Each dimension
of $M$ corresponds to a different measurement result, $m$.
The operation is specified by:
\begin{eqnarray}
\label{eqtn: equivmap}
{\cal M}(\rho) = \sum_m {\cal N}_m(\rho) \otimes |m\rangle \langle m|,
\end{eqnarray}
where $|m\rangle$ is some set of orthogonal states corresponding to the
measurement results which may occur.
This map is an ``all-quantum'' version of the observed channel.

Since our upper bounds to the capacity of an unobserved
channel apply also to channels with output Hilbert spaces of different
dimensionality than the input space, they apply to this map as well.
It is easily verified that the coherent information for the map
${\cal M}$ acting on $\rho$ is the same as the average coherent information
for the observed channel
${\cal N}_m$ acting on $\rho$, which appears in
(\ref{eqtn: fidelity lemma for general operations}) and in the bound
(\ref{eqtn: observed capacity}).  To show this, define
\begin{eqnarray}
p_m \equiv {\rm tr} ({\cal N}_m(\rho^Q)),
\end{eqnarray}
where we are again working in the $RQ$ picture of operations.
Then
$\rho^{Q'} = {\cal M}(\rho^Q)$ is given by (\ref{eqtn: equivmap}),
so that 
\begin{eqnarray}
S(\rho^{Q'}) = H(p_m) + \sum_m p_m S(\frac{{\cal N}_m (\rho^Q)}{p_m})
\end{eqnarray} since the density matrices
${\cal N}_m (\rho^{Q}) \otimes |m \rangle \langle m|$ are mutually orthogonal.  
Also, 
\begin{eqnarray}
\rho^{R'Q'} = ({\cal I} \otimes \sum_m {\cal N}_m^*)(\rho^{RQ}),
\end{eqnarray}
 where by definition ${\cal N}_m^*(\rho) = 
{\cal N}_m (\rho) \otimes |m \rangle \langle m|.$  Thus
\begin{eqnarray}
S(\rho^{R'Q'}) = H(p_m) + \sum_m p_m S(\frac{({\cal I} \otimes
{\cal N}_m) (\rho^{RQ})}{p_m}).
\end{eqnarray}
Hence the coherent information is
\begin{eqnarray}
I(\rho^Q,{\cal M}) & = &  \sum_m p_m [S(\frac{{\cal N}_m (\rho^Q)}{p_m}) 
- S(\frac{({\cal I} \otimes
{\cal N}_m) (\rho^{RQ})}{p_m})], \nonumber \\
& & 
\end{eqnarray}
which can be rewritten as the average coherent information for $\{{\cal N}_m\},$
\begin{eqnarray}
I(\rho^Q,{\cal M}) & = &  \sum_m p_m I(\rho^Q, {\cal N}_m).
\end{eqnarray}
So an application of the bound (\ref{eqtn: general bound}) 
on the rate of transmission
through the unobserved channel ${\cal M}$ shows the expression 
on the right hand side of (\ref{eqtn: observed capacity}) which bounds
the capacity of the observed channel $\{{\cal N}_m\}$ also bounds
the capacity of ${\cal M}$. 
This result provides some evidence
for the 
intuitively reasonable proposition that 
${\cal M}$ and $\{ {\cal N}_m\}$ are equivalent with respect to 
transmission of quantum information.

Bennett et. al \cite{Bennett97a} derive capacities for three simple
channels which may be viewed as taking the form (\ref{eqtn: equivmap}).
The {\em quantum erasure channel} takes the input state to a fixed state orthogonal to the input state with probability $\epsilon$;  otherwise, it transmits the state undisturbed.  An equivalent observed channel would with probability $\epsilon$ replace the
input state with a standard pure state $|0\rangle \langle 0|$ within
the input subspace, and also provide classical information as to whether
this replacement has occurred or not. The {\em phase erasure channel} randomizes the phase of a qubit with probability $\delta$,
and otherwise transmits the state undisturbed;  it also
supplies classical information as to which of these alternatives
occurred.  The {\em mixed erasure/phase-erasure channel} may either erase
or phase-erase, with exclusive probabilities $\epsilon$ and $\delta$.
Bennett et. al. note that the capacity $\max ( 0,1 - 2 \epsilon )$
of the erasure channel
is in fact the one-shot maximal coherent information.  We have 
verified that the capacities they derive for the phase-erasure
channel ($1-\delta$) and the mixed erasure/phase-erasure channel 
$\max ( 0, 1 - 2\epsilon - \delta )$ are the same
as the one-shot maximal average coherent information for the corresponding 
observed channels, lending some additional support to the view
that the bounds we have derived here are in fact the capacities.

\subsection{Relationship to Unobserved Channel}
\label{subsect: relationship to unobserved channel}

Suppose a quantum system passes through a
channel, interacts with an environment, and then measurements are performed
on the {\em environment alone}. How is the capacity of this observed channel
related to the capacity of the channel which results if {\em no measurement}
had been performed on the environment? Physically, it is clear that the capacity
when measurements are performed must be at least as great as
when no measurements on the environment are performed, since the
decoder can always ignore the result of the measurement. In this subsection
we show that bounds we have derived on channel capacity have this same property:
observation of the environment can never decrease the bounds we have obtained.

Suppose $\{ {\cal N}_m \}$ are the operations describing the different possible
measurement outcomes. Then the operation describing the same channel, but
without any observation of the environment, is
\begin{eqnarray}
{\cal N} = \sum_m {\cal N}_M. \end{eqnarray}

Recall the expressions for the bound on the capacity of the unobserved channel,
\begin{eqnarray}
C({\cal N}) = \lim_{n \rightarrow \infty}
	 \max_{{\cal C}^n,\rho} \frac{I(\rho,{\cal N}^{\otimes n} 
	\circ {\cal C}^n)}{n},
\end{eqnarray}
and the observed channel,
\begin{eqnarray}
C(\{ {\cal N}_m \}) = \lim_{n \rightarrow \infty} \max_{{\cal C}^n, \rho} \nonumber \\
\sum_{m_1,...m_n} \mbox{tr}(({\cal N}_{m_1} \otimes \cdots \otimes {\cal N}_{m_n}
 \circ {\cal C}^n )(\rho)) \nonumber \\
\frac{I(\rho, {\cal N}_{m_1} \otimes \cdots \otimes {\cal N}_{m_n}\circ {\cal C}^n)}{n},
\end{eqnarray}
But the generalized convexity theorem (\ref{eqtn: abstract subadditivity})
for coherent information implies that
\begin{eqnarray}
	\sum_{m_1,...,m_n} \mbox{tr}(( 
	{\cal N}_{m_1}\otimes \cdots \otimes {\cal N}_{m_n}
\circ {\cal C}^n)(\rho)) \nonumber \\
\frac{I(\rho,{\cal N}_{m_1}\otimes \cdots \otimes {\cal N}_{m_n} \circ {\cal C}^n)}{n} \nonumber \\
	\leq
    \frac{I(\rho,{\cal N}^{\otimes n} \circ {\cal C}^n)}{n}, 
\end{eqnarray}
and thus
\begin{eqnarray}
C({\cal N}) \leq C(\{ {\cal N}_m \}). 
\end{eqnarray}

To see that this inequality may sometimes be strict, return to the
example considered earlier in this section
in the context of teleportation. In that case it
is not difficult to verify that
\begin{eqnarray}
0 = C({\cal N}) < C(\{ {\cal N}_m \}) = 1. 
\end{eqnarray}

What these results show is that our bounds on the channel capacity are never
made any worse by observing the environment, but sometimes they can be
made considerably better. This is a property that we certainly expect the
quantum channel capacity to have, and we take as an encouraging sign that
the bounds we have proved in this paper are in fact achievable, that is,
the true capacities.

\subsection{Discussion}
\label{subsect: observed discussion}

All the questions asked about the bounds on channel capacity for an
unobserved channel can be asked again for the observed channel: questions
about achievability of bounds, the differences in power achievable by
different classes of encodings and
decodings, and so on. We do not address these problems here, beyond
noting that they are important problems which need to be
addressed by future research.

Many new twists on the problem of the quantum noisy channel arise when an
observer of the environment is allowed. For example, one might consider the
situation where the classical channel connecting the observer to the
decoder is noisy. What then are the resources required to transmit
coherent quantum information?

It may also be interesting to prove results relating the
classical and quantum resources that are required to perform a certain
task. For example, in teleportation it can be shown that one requires
not only the quantum channel, but also two bits of classical information,
in order to transmit quantum information with perfect
reliability.

\section{Conclusion}
\label{sect: conc}

In this paper we have shown that different information transmission
problems may result in different channel capacities for the same noisy
quantum channel.
We have developed some general techniques for proving upper bounds on
the amount of information that may be transmitted reliably through
a noisy quantum channel.

Perhaps the most interesting thing about the quantum noisy channel problem
is to discover what is new and essentially {\em quantum} about the
problem. The following list summarizes what we believe are the
essentially new features:
\begin{enumerate}

\item The insight that there are many essentially different information
transmission problems in quantum mechanics, all of them of interest
depending on the application. These span a spectrum between two extremes:

\begin{itemize}

\item The transmission of a discrete set of mutually orthogonal 
quantum states through the
channel. Such problems are problems of transmitting
classical information through a noisy quantum channel.

\item The transmission of entire subspaces of quantum states
through the channel, which necessarily keeps all other quantum
resources, including entanglement, intact.
This is likely to be of interest in applications such as
quantum computation, cryptography and teleportation where
superpositions of quantum states are crucial. Such problems
are problems of transmitting 
coherent quantum information through a noisy quantum channel.

\end{itemize}

Both these cases and a variety of intermediate cases 
are important for specific applications. For
each case, there is great interest in considering
different classes of allowed encodings and decodings. For example,
it may be that encoding and decoding can only be done using local
operations and one-way classical communication. This may give rise to
a different channel capacity than occurs if we allow non-local
encoding and decoding. Thus there are different noisy channel problems
depending on what classes of encodings and decodings are allowed.

\item The use of quantum entanglement to construct examples where
the quantum analogue of the classical pipelining inequality
$H(X:Z) \leq H(Y:Z)$ for a Markov process
$X \rightarrow Y \rightarrow Z$, fails to hold (cf. eqtn.
(\ref{eqtn: example 1})).

\item The use of quantum entanglement to construct examples where the
subadditivity property of mutual information,
\begin{eqnarray}
H(X_1,\ldots,X_n : Y_1,\ldots,Y_n) & \leq & \sum_i H(X_i : Y_i), \nonumber \\
 & &
\end{eqnarray}
fails to hold (cf. eqtn. (\ref{eqtn: example 2})).

\end{enumerate}

There are many more interesting open
problems associated with the noisy channel
problem than have been addressed here.
The following is a sample of those problems which we believe to be
particularly important:
\begin{enumerate}

\item The development of an effective procedure for determining
channel capacities. We believe that this is the most important
problem remaining to be addressed. Assuming our upper bound
\begin{eqnarray}
C({\cal N}) = \lim_{n \rightarrow \infty} \max_{\rho,{\cal C}}
	I(\rho, {\cal N}^{\otimes n} \circ {\cal C})
\end{eqnarray}
is, in fact, the channel capacity for general encodings,
it still remains to find an effective procedure for evaluating
this quantity. Both maximizations can be done relatively easily,
since they are of a continuous function over a compact set.
However, we do not yet understand the convergence of the limit
well enough to have an effective procedure for evaluating
this quantity.

\item Once an effective procedure has been obtained for evaluating
channel capacities, it still remains to develop {\em good}
numerical algorithms for performing the evaluation. Assuming
that the evaluation involves some kind of maximization of
the coherent information, it becomes important to know whether
the coherent information is concave in $\rho.$
Combined with the convexity of the coherent information
in the operation this would give a powerful tool for the development
of numerical algorithms for the determination of channel capacity.  

\item Estimation of channel capacities for realistic channels. This
work could certainly be done theoretically and perhaps also
experimentally. 
Recent work on {\em quantum process tomography}
\cite{Poyatos97a,Chuang96a} points the way toward experimental
determination of the quantum channel capacity. A
related problem is to analyze how stable the determination of 
channel capacities is with respect to experimental error.

\item As suggested in
subsection \ref{subsect: other encoding protocols}
it would be interesting to see what channel capacities are
attainable for different classes of allowable encodings and / or
decodings, for example,
encodings where the encoder is only allowed to do local operations
and one-way classical communication, or encodings where the encoder
is allowed to do local operations and two-way classical communication.
We have showed how to prove bounds on the channel capacity in these
cases; whether these bounds are attainable is
unknown.

\item The development of rigorous general techniques
for proving attainability
of channel capacities, which may be applied to different classes of allowed
encodings and decodings.

\item Finding the capacity of a noisy quantum channel for classical
information. A related problem arises in the context of
{\em superdense coding}, where one half of an EPR pair can be used
to send two bits of classical information. It would be interesting to
know to what extent this performance is degraded if the pair of qubits
shard between sender and receiver is not an EPR pair, but rather the
sharing is done using a noisy quantum channel, leading to a decrease
in the number of classical bits that can be sent. Given a noisy
quantum channel, what is the maximum amount
of classical information that can be sent in this way?

\item All work done thus far has been for discrete channels,
that is, channels with finite dimensional state
spaces. It is an important and non-trivial
problem to extend these results
to channels with infinite dimensional state spaces.

\item A more thorough study of noisy channels which have a classical
side channel. Can the classical information obtained by an observer
be related to changes in the channel capacity? What if the classical
side channel is noisy? Many other fascinating problems, too many
to enumerate here, suggest themselves in this context.

\end{enumerate}

There are many other ways the classical results on noisy channels
have been extended - considering channels with
{\em feedback}, developing
{\em rate-distortion} theory, understanding {\em networks} consisting
of more than one channel, and so on. Each of these could give rise
to highly interesting work on noisy quantum channels. It is also to be
expected that interesting new questions will arise as experimental
efforts in the field of quantum information develop further.
Perhaps of
chief interest to us is to develop a still clearer understanding of
the essential differences between the quantum noisy channel and
the classical noisy channel problem.

\section*{Acknowledgments}

We thank Carlton~M.~Caves, Isaac~L.~Chuang, Richard~Cleve,
David~P.~DiVincenzo, Christopher~A.~Fuchs, E.~Knill, and John~Preskill for many
instructive and enjoyable discussions about quantum information. This
work was supported in 
part by the Office of Naval Research (Grant No.\ N00014-93-1-0116).
We thank the 
Institute for Theoretical Physics for its hospitality and for the 
support of the National Science Foundation (Grant No.\ PHY94-07194).  
MAN acknowledges financial support from the Australian-American Educational 
Foundation (Fulbright Commission).

\appendix

\section{Existence of limits}
\label{appendix: limit lemma}

This appendix contains a lemma that can be used to
prove the existence of several limits that appear in this paper.

{\em Lemma}: Suppose $c_1,c_2,\ldots$ is a nonnegative  sequence such that $c_n \leq kn$
for some $k \geq 0$, and
\begin{eqnarray} \label{eqtn: abstract superadditivity}
c_m + c_n \leq c_{m+n}, \end{eqnarray}
for all $m$ and $n$. Then
\begin{eqnarray} \label{eqtn: lim exists}
\lim_{n \rightarrow \infty} \frac{c_n}{n} \end{eqnarray}
exists and is finite.

{\em Proof}

Define
\begin{eqnarray}
c \equiv \limsup_n \frac{c_n}{n}. \end{eqnarray}
This always exists and is finite, since $c_n \leq kn$ for some $k \geq 0$.
Fix $\epsilon > 0$ and choose $n$ sufficiently large that
\begin{eqnarray}
\frac{c_n}{n} > c - \epsilon. \end{eqnarray}
Suppose $m$ is any integer strictly greater than $\max(n,n/\epsilon)$.
Then by (\ref{eqtn: abstract superadditivity}),
\begin{eqnarray} \label{eqtn: intermediate appendix}
\frac{c_m}{m} \geq \frac{c_n}{n} \frac{n}{m} \left( 1 + \frac{c_{m-n}}{c_n}
	\right). \end{eqnarray}
Using the fact that $l c_n \leq c_{ln},$ (an immediate consequence 
of (\ref{eqtn: abstract superadditivity})) with $l = \lfloor 
\frac{m}{n} \rfloor- 1$ gives
\begin{eqnarray}
\frac{c_{m-n}}{c_n} & \geq & \lfloor \frac{m}{n} \rfloor - 1 \\
 & \geq & \frac{m}{n} - 2, \end{eqnarray}
where $\lfloor x \rfloor$ is the integer immediately below $x$. Plugging the
last inequality into (\ref{eqtn: intermediate appendix}) gives
\begin{eqnarray}
\frac{c_m}{m} \geq \frac{c_n}{n} \left( 1 - \frac{n}{m} \right).
\end{eqnarray}
But $-n/m > -\epsilon$ and $c_n/n \geq c - \epsilon$, so
\begin{eqnarray}
\frac{c_m}{m} \geq (c-\epsilon)(1-\epsilon). \end{eqnarray}
This equation holds for all sufficiently large $m$, and thus
\begin{eqnarray}
\liminf_n \frac{c_n}{n} \geq (c-\epsilon)(1-\epsilon). \end{eqnarray}
But $\epsilon$ was an arbitrary number greater than $0$, so letting
$\epsilon \rightarrow 0$ we see that
\begin{eqnarray}
\liminf_n \frac{c_n}{n} \geq c = \limsup_n \frac{c_n}{n}. \end{eqnarray}
It follows that $\lim_n c_n/n$ exists, as claimed.

\section{Maxima of the Coherent Information}
\label{appendix: convexity}
Various convexity and concavity properties are useful in calculating
classical channel capacities, and the same is true in the quantum
situation. This appendix is devoted to an explication of the basic
properties of convexity and concavity related to the coherent
information and the relation of these properties to expressions such
as (\ref{eqtn: general bound}).
  
A {\em convex set}, $S$, is a subset of a vector space such that
given any two points $s_1, s_2 \in S$ and any
$\lambda$ such that $0 < \lambda < 1$, then the {\em convex combination},
$\lambda s_1 + (1-\lambda) s_2$, is also an element of $S$.
Geometrically, this means that given any
two points in the set, the line joining them is also in the set. An
{\em extremal point} of $S$ is a point $s$ which cannot be formed from the
convex combination of any other two points in the set. A
{\em convex function} $f$ on $S$ is a real-valued function such that
for any $\lambda$ satisfying $0 < \lambda < 1$,
\begin{eqnarray}
f(\lambda s_1 + (1-\lambda) s_2) \le \lambda f(s_1) + (1-\lambda) f(s_2); 
\end{eqnarray}
a concave function satisfies the same condition but with the
inequality reversed.

The first useful fact about maxima is the following:

{\em Local maximum is a global maximum}:
Suppose $f$ is a concave function on a convex set $S$. Then a local maximum
of $f$ is also a global maximum of $f$.

This follows by supposing that $s_1$ and $s_2$ are distinct local
maxima. If $f(s_1) < f(s_2)$, say, then
\begin{eqnarray}
f(\lambda s_1 + (1-\lambda) s_2) & \geq & \lambda f(s_1) + (1-\lambda)f(s_2) \\
 & > & f(s_1), \end{eqnarray}
by concavity of $f$. By choosing sufficiently small values of $\lambda$
we see that this violates the fact that $s_1$ is a local maximum.
Thus $f$ has the same value for all local
maxima, from which it follows that all local maxima are also global
maxima for the function.  If the coherent information turns out
to be concave in the input density operator, this property will be
useful in evaluating capacity bounds such as (\ref{eqtn: general bound}).

The following lemma, from \cite{Marcus92}, is extremely useful in computing the
maxima of convex functions on convex sets.

{\em Convexity Lemma:}
Suppose $f$ is a continuous convex function on a compact,
convex set, $S$. Then there is an extremal point at which $f$
attains its global maximum.

The proof is obvious. The reason for our interest in the
proof is because for fixed $\rho$ and
trace-preserving operations ${\cal E}$, the coherent information
$I(\rho,{\cal E})$ is a convex, continuous function of the operation
${\cal E}$. The set of trace-preserving quantum operations forms a compact,
convex set, and thus by the convexity lemma,
$I(\rho,{\cal E})$ attains its maximum for a quantum operation
${\cal E}$ which is extremal in the set of all trace-preserving quantum
operations.

Choi \cite{Choi75a} has proved that any extremal point in the
set of trace-preserving quantum operations has a set of operation elements
$\{ A_i \}$ such that

\begin{enumerate}

\item There are at most $d$ elements $A_i$. This is to be contrasted with the
general situation, where there may be up to $d^2$ elements.

\item The $A_i$ are linearly independent.

\end{enumerate}

This result provides a considerable saving in the class of quantum operations
that must be optimized over in order to numerically calculate
expressions of the form (\ref{eqtn: general bound}). Unfortunately, this
only takes us part of the way towards proving that the
expressions (\ref{eqtn: general bound}) and (\ref{eqtn: unitary bound})
are identically equal, or, alternatively, it suggests
a starting point for a search for counterexamples to the proposition
that the two quantities are equal. If the extremal points of the set of quantum
operations were the unitary operations we would be done. However that
is not the case, as the above theorem shows.

\end{multicols}


\begin{thebibliography}{10}

\bibitem{Shannon48a}
C.~E. Shannon, Bell System Tech. J. {\bf 27},  379  (1948).

\bibitem{Shannon49a}
C.~E. Shannon and W. Weaver, {\em The Mathematical Theory of Communication}
  (University of Illinois Press, Urbana, 1949).

\bibitem{Cover91a}
T.~M. Cover and J.~A. Thomas, {\em Elements of Information Theory} (John Wiley
  and Sons, New York, 1991).

\bibitem{Schumacher96a}
B.~W. Schumacher, Phys. Rev. A {\bf 54},  2614  (1996).

\bibitem{Schumacher96b}
B.~W. Schumacher and M.~A. Nielsen, Phys. Rev. A {\bf 54},  2629  (1996).

\bibitem{Bennett96a}
C.~H. Bennett, D.~P. DiVincenzo, J.~A. Smolin, and W.~K. Wootters, Phys. Rev. A
  {\bf 54},  3824  (1996).

\bibitem{Lloyd97a}
S. Lloyd, LANL e-print quant-ph/9604015 (1996).

\bibitem{Bennett97a}
C.~H. Bennett, D. DiVincenzo, and J. Smolin, LANL e-print quant-ph/9701015
  (1997).

\bibitem{Bennett93a}
C.~H. Bennett {\it et~al.}, Phys. Rev. Lett. {\bf 70},  1895  (1993).

\bibitem{Nielsen96b}
M.~A. Nielsen and C.~M. Caves, LANL e-print quant-ph/9608001  (1996).

\bibitem{Kraus83a}
K. Kraus, {\em States, Effects, and Operations} (Springer-Verlag, Berlin,
  1983).

\bibitem{Hellwig70}
K. Hellwig and K. Kraus, Commun. Math. Phys. {\bf 16},  142  (1970).

\bibitem{Choi75a}
M.-D. Choi, Linear Algebra and Its Applications {\bf 10},  285  (1975).

\bibitem{Nielsen97a}
M.~A. Nielsen, H. Barnum, C.~M. Caves, and B.~W. Schumacher, unpublished
  (1997).

\bibitem{Peres93a}
A. Peres, {\em Quantum Theory: Concepts and Methods} (Kluwer Academic,
  Dordrecht, 1993).

\bibitem{Cohen-Tannoudji77a}
C. Cohen-Tannoudji, B. Diu, and F. Lalo\"e, {\em Quantum Mechanics} (John Wiley
  and Sons, New York, 1977).

\bibitem{Hughes89}
R.~I.~G. Hughes, {\em The Structure and Interpretation of Quantum Mechanics}
  (Harvard University Press, Cambridge, 1989).

\bibitem{Luders51}
G. L{\"u}ders, Annalen der Physik {\bf 8},  323  (1951).

\bibitem{Jozsa95a}
R. Jozsa, J. Mod. Optics {\bf 41},  2315  (1995).

\bibitem{Knill97a}
E. Knill and R. Laflamme, Phys. Rev. A {\bf 55},  900  (1997).

\bibitem{Nielsen96c}
M.~A. Nielsen, LANL e-print quant-ph/9606012  (1996).

\bibitem{Wehrl78a}
A. Wehrl, Rev. Mod. Phys. {\bf 50},  221  (1978).

\bibitem{Shor96a}
P.~W. Shor and J.~A. Smolin, LANL e-print quant-ph/9604006  (1996).

\bibitem{Schumacher95}
B. Schumacher, Phys. Rev. A {\bf 51},  2738  (1995).

\bibitem{Jozsa94a}
R. Jozsa and B. Schumacher, J. Mod. Optics {\bf 41},  2343  (1994).

\bibitem{Barnum96}
H. Barnum, C.~A. Fuchs, R. Jozsa, and B. Schumacher, Phys. Rev. A {\bf 54},
  4707  (1996).

\bibitem{Poyatos97a}
J.~F. Poyatos, J.~I. Cirac, and P. Zoller, Phys. Rev. Lett. {\bf 78},  390
  (1997).

\bibitem{Chuang96a}
I.~L. Chuang and M.~A. Nielsen, LANL e-print quant-ph/9610001  (1996).

\bibitem{Marcus92}
M. Marcus and H. Minc, {\em A Survey of Matrix Theory and Matrix Inequalities}
  (Dover, New York, 1992).

\end{thebibliography}
\end{document}